\documentclass{iopart}

\usepackage{braket}
\usepackage{iopams, setstack}
\usepackage{mathbbol, bm, paralist, graphicx, placeins}

\usepackage{pifont,color}
\definecolor{darkblue}{rgb}{.1,0,.3}
\usepackage[dvipdfm,breaklinks,colorlinks,bookmarksopen,citecolor=darkblue,linkcolor=darkblue,urlcolor=darkblue]{hyperref}

\usepackage{footmisc}
\usepackage{perpage}
\MakePerPage[2]{footnote}

\newcommand{\on}[1]{\mathop{\mbox{#1}}}
\renewcommand{\text}[1]{\mbox{#1}}
\newcommand{\bEq}{\begin{equation}}
\newcommand{\eEq}{\end{equation}}
\renewcommand{\phi}{\varphi}
\newcommand{\wrt}{w.\,r.\,t.\ }
\newcommand{\req}[1]{Eq.\ (\ref{#1})}
\newcommand{\ReIm}{{\left\{\rm \scriptstyle Re \atop \scriptstyle Im \right\}}}
\newcommand{\OneI}{{\left\{1\atop\rmi\right\}}}

\newcommand{\PlusMinus}{{\left\{+\atop -\right\}}}

\begin{document}

\title{A variational method based on weighted graph states}

\author{Simon Anders$\,^1$, Hans J.\ Briegel$\,^{1,2}$, Wolfgang D\"ur$\,^{1,2}$}

\address{$^1$ Institut f\"ur Theoretische Physik, Universit\"at Innsbruck, Technikerstra\ss{}e~25, 6020~Innsbruck, Austria}
\address{$^2$ Institut f\"ur Quantenoptik und Quanteninformation der \"OAW, Technikerstra\ss{}e~21a, 6020~Innsbruck, Austria}
\ead{sanders@fs.tum.de \textrm{(S.\,An.)}}

\begin{abstract}
In a recent article [\href{http://dx.doi.org/10.1103/PhysRevLett.97.107206}{Phys.\ Rev.\ Lett.\ \textbf{97} (2006), 107206}]\nocite{WGSMinPRL}, we have presented a class of states which is suitable as a variational set to find ground states in spin systems of arbitrary spatial dimension and with long-range entanglement. Here, we continue the exposition of our technique, extend from spin 1/2 to higher spins and use the boson Hubbard model as a non-trivial example to demonstrate our scheme. 
\end{abstract}

\pacs{02.70.c, 05.30.Jp, 03.67.Mn, 75.10.Jm, 75.40.Mg\\[2em] {3 June 2007}}

\maketitle


\section{Introduction}

Spins or harmonic oscillators on a lattice form a class of models which have been studied intensively in statistical physics. Understanding them is the key to many problems in condensed matter systems, especially regarding magnetic phenomena but also electrical and heat conduction and many other aspects. As the importance of quantum phase transitions \cite{Sachdev_book, QPT_review_Vojta} has been more and more realized, interest in the ground states of \textit{quantum} spin models grew. While the relevance of entanglement for quantum phase transitions was initially not fully appreciated, it is now a vivid area of research (e.\,g., \cite{entanglement_at_QPT, Entanglement_in_Quant_Crit}), and many researchers feel that paying explicit attention to entanglement features is vital for further progress in numerical methods for the treatment of spin models \cite{entanglement_and_DMRG,DMRG_seen_from_QIT,Latorre_on_MPS_and_QuantSim}. Although quantum phase transitions nominally only occur at zero temperature, their presence has great influence on the system properties at finite temperature, namely leading to the break-down of quasiparticle descriptions. Hence, studying the ground state of spin models holds promises to understand experimentally observed features of such systems, not the least of which is high-temperature superconductivity. Finally, spin models (including bosons on a lattice) are an ideal way to model optical lattices, which are currently researched with exciting successes in theory and experiment (reviewed in \cite{Review_Optical_Lattices}).

While there are some exactly solvable spin models in one spatial dimension \cite{1D_exact_models}, for nearly all models in higher dimensions approximative techniques have to be used. A variety of quite different techniques have been developed: Most prominently, these are quantum Monte Carlo techniques, where recent progress has been acieved especially in the context of the so-called world-line Monte Carlo methods (\cite{LoopQMC_first,WormQMC_first}, reviewed in \cite{QMC_Lattice_Review_04}). For one-dimensional systems, extraordinary accuracy has become possible with the density matrix renormalisation group (DMRG) algorithm (\cite{White_DMRG_1,White_DMRG_2}, review \cite{Schollwoeck_Review}). Recently, this algorithm was extended to allow the calculation of not only ground state properties but also of thermal states \cite{DMRG_thermal_B,DMRG_thermal} and time evolutions \cite{tDMRG_first,tDMRG,VidalTEvol,time-evolution_DMRG_2}. Also, an extension to higher spatial dimensions has been proposed \cite{Verstr_Renorm}. Its usability in practice has been demonstrated only very recently \cite{PEPS_for_BH}.

All these variations of DMRG are based on the same class of variational\footnote{Strictly speaking, only the fixed-length phase of DMRG can be called a variational method.} states, namely matrix product states  \cite{DMRG_uses_MPS}. We have recently found \cite{WGSMinPRL} that another class of states, namely the so-called weighted graph states (WGS), first studied in different context in \cite{spin_chain_long_range,spin_gases}, is also quite promising as ansatz for variational approximation of ground states of spin systems. Its particular advantage is the unlimited amount of entanglement that can be present. Hence, we consider our technique as especially promising for systems with long-range entanglement such as critical systems. A further key difference of our states to matrix product states is that their mathematical structure does not reflect any spatial geometry (while the product of matrices in a matrix product state reflects a chain or ring geometry, as studied especially in \cite{DMRG_uses_MPS,DMRG_seen_from_QIT}) and hence may be expected to be equally suitable for higher dimensions (2D or 3D) as for 1D. Hence, even though we probably cannot compete with the astounding accuracy of DMRG in 1D, we aim to provide a complementary alternative to the higher-dimension generalisations of DMRG \cite{Verstr_Renorm}. In \cite{WGSMinPRL}, we presented this technique and demonstrated its use for simple spin-1/2 systems in one and two dimensions. In the present article we explain our method in much more detail, show new results we have obtained since then (especially regarding the treatment of spins higher than spin-1/2, and concerning heuristics to perform the minimizations) and tests its usefuleness on practical examples. The article is self-contained and does not assume the reader's familiarity with weighted graph states or the content of \cite{WGSMinPRL}.

This article is organised as follows: We start in Sec.\ \ref{var_gen} by reviewing some general observations about variational methods. In Sec.\ \ref{our_class} we describe our class of variational states as a generalisation of weighted graph states and discuss their parametrisation. Section \ref{calc_rdm} explains how reduced density matrices of these states are calculated in an efficient manner in order to be able to evaluate expectation values of observables, including energy. To test our method, we show results for calculations on two different models (namely the XY model and the Bose-Hubbard model) in Sec. \ref{models}. In a variational method, a crucial part is finding a state within the given class that minimises the energy as well as possible. Our techniques for doing so are the topic of Sec.\ \ref{perf_min}. We add some further notes on the details of our numerical implementation and its performance (Sec. \ref{implementation}), and finish with a conclusion and an outlook on further work (Sec.\ \ref{conclusion_and_outlook})

\section{General considerations on variation} \label{var_gen}

For a Hamiltonian $H$ that is too large to diagonalise one can approximate the ground state using the Rayleigh-Ritz variational method. One uses a family of states $\Ket{\Psi(\mathbf{x})}$ which depend on some parameter $\mathbf{x}$. It may be better to see this as a map from a parameter space $\mathbb{R}^K$ to a Hilbert space $\mathcal{H}$:
\[\Psi:\, \mathbb{R}^K \rightarrow \mathcal{H},\quad \mathbf{x} \mapsto \Ket{\Psi(\mathbf{x})}.\]

One then solves the minimisation problem
\bEq E_{\rm min} = \on{min}_{\mathbf{x}\in\mathbb{R}^K} E(\mathbf{x});\quad \text{ with } E(\mathbf{x}) = \frac {\Braket{\Psi({\mathbf{x})} | H | \Psi({\mathbf{x}})}} {\Braket{\Psi({\mathbf{x}}) | \Psi({\mathbf{x}})}} \label{minprob} \eEq
in order to obtain an upper bound $E_{\rm min}$ to the ground state energy and an approximation $\Ket{\Psi(\mathbf{x_{\rm min}})}$ for the ground state.

For this to give good results, the map $\Psi$ has to fulfil the following conditions:

(i) There must be an efficient algorithm to calculate the expectation value of observables for any state $\Psi(\mathbf{x})$. In principle, it is sufficient to be able to calculate $\Braket{\Psi({\mathbf{x})} | H | \Psi({\mathbf{x}})}$ and $\Braket{\Psi({\mathbf{x}}) | \Psi({\mathbf{x}})}$, but if one wants not only to bound the ground state energy, but also analyse the ground state approximant $\Ket{\Psi(\mathbf{x_{\rm min}})}$, it is desirable to be able to calculate expectation values for other observables, too.

``Efficient'' means here a computation time at most polynomial in the number of parameters $K$. As the dimension of the Hilbert space $\mathcal{H}$ typically scales exponentially in the number $N$ of constituents of the system, we  want $K$ to \textit{not} do the same. Thus, the map $\Psi$, considered as a a family $\Psi_N$ of maps for different system sizes $N$, should be such that the dimension $K$ of its domain scales only polynomially with $N$, and thus logarithmically with $\on{dim} \mathcal{H}$.

(ii) There should be reason to expect that there are states $\Ket{\Psi(\mathbf{x})}$ within the range of the map $\Psi$ that have large overlap with the true ground state or at least an energy near to the true ground state energy. As the range of $\Psi$ is a sub-manifold of $\mathcal{H}$ of dimension at most $K\ll \dim\mathcal{H}$, this requires it either to be folded and twisted in a quite peculiar way to reach many different regions of $\mathcal{H}$, or to happen to occupy the same small part of $\mathcal{H}$ as the ground state.  Typically, it is not possible to prove such a statement, and one hence has to do with heuristic arguments or numerical evidence.

(iii) There should be reason to expect that the minimisation programme (\ref{minprob}) succeeds in finding a good minimum and does not get stuck in a bad local minimum. It is often not justified to hope to find the global minimum, but a local minimum of an energy only slightly higher than that of the global minimum is hardly worse.

Whether the minimisation can succeed depends on the ``energy landscape'', i.\,e. the graph of $E(\mathbf{x})$. If this landscape has many local minima, a na\"ive multi-start optimisation cannot succeed. Often, the number of local minima increases exponentially with $N$ or $K$, which may render a method that is efficient for small systems useless for larger ones. Hence, one usually has to succeed in tailoring a heuristics that helps to find good minima for the specific kind of energy landscape one has to deal with.

One of the best studied variational methods is finite-length DMRG, and we shall illustrate the conditions given above by briefly discussing how DMRG (in the formulation of Ref. \cite{DMRG_seen_from_QIT}) fulfils them. For DMRG, the class of variational states are the matrix product states \cite{DMRG-MPS-PRL,DMRG_uses_MPS}. For an $N$-site matrix product state, an efficient algorithm exists to evaluate the expectation value of any observable that can be written as a sum of tensor products of local operators in time linear in $N$. This meets condition (i). The expectation that a matrix product state is a good approximant for the ground state of a generic 1D system (condition (ii)) is the very rationale that led White and Noack to their idea of keeping the lowest-lying eigenstates not of the short-range Hamiltonian but of the corresponding density matrix as explained e.\,g. in \cite{Whites_DMRG_review}. The fact that condition (iii) is fulfilled, i.\,e. that the ``sweeping procedure'' of finite-length DMRG does not get stuck in local minima is somewhat mysterious, especially in the light of the possibility of construction of Hamiltonian for which this cannot be avoided \cite{DMRG_Minimization_is_NP_complete}. Nevertheless, the construction principle, as exposed in \cite{DMRG_seen_from_QIT}, shows that the matrix for each site has direct influence only on this site and its neighbours, i.\,e. matrix product states allow for an essentially local description of states despite the existence of significant amount of entanglement. Hence, is seems natural that ---barring ``pathological'' cases such as those discussed in \cite{DMRG_Minimization_is_NP_complete}--- the local variation of matrices during sweeps allows for a good minimisation, provided the initial $N$-site state was chosen well (which is the task of the so-called ``warm-up'', which uses infinite-length DMRG). Furthermore, Wolf et al.\ have recently shown a close connection between approximability and R\'enyi entropy for matrix product states \cite{MPS_and_Renyi}.

We shall come back to some of these points when comparing our variational states with matrix product states at the end of Sec.\ \ref{our_class}.

\section{The class of variational states} \label{our_class}

\subsection{Basic idea}

Our class of quantum states derives from the so-called weighted graph states, which were introduced in \cite{spin_chain_long_range,Long1QC} and also used in \cite{spin_gases}. They are a generalisation of graph states (introduced in \cite{cluster_states}, see \cite{marcs_review} for a review). For a Hilbert space of $N$ qubits, they are defined as\footnote{In this article, superscripts in parentheses always indicate the spins an operator acts on. Hence $W_\phi$ is an operator defined on a 2-spin space, while $W_\phi^{(ab)}$ is defined on the full $N$-spin Hilbert space, but has support only on spins $a$ and $b$.} 
\bEq \Ket{\Gamma} = \prod_{a=1}^N \prod_{b=a+1}^N W_{\phi_{ab}}^{(ab)} \Ket{+}^{\otimes N}, \label{simpleWGS} \eEq
where a product of \textit{phase gates} $W_\phi$ is applied onto a tensor product of $\Ket{+}=(\Ket{0}+\Ket{1})/\sqrt{2}$ states. These phase gates are two-qubit operation, diagonal in the computational basis, and of the form \footnote{In \cite{spin_chain_long_range,spin_gases}, the notation $U_{\phi_{ab}}$ is used instead of $W_{\phi_{ab}}^{(ab)}$. Here, we use the $W$ to emphasise that it is a specific, and not some general unitary. Note also the absence of a minus sign in the exponential $\rme^{\rmi\phi}$, which differs from the convention used in \cite{spin_chain_long_range}.}
\bEq W_\phi = \on{diag} (1,1,1,\rme^{\rmi\phi}) = \exp \left[i \frac{\phi}{2} \left(\mathbb{1}-\sigma_z\right)\otimes \left(\mathbb{1}-\sigma_z\right)\right]. \label{phase_gate_1_2} \eEq

It may help to see the effect of $W$ on small states. This is, e\,g., a three-qubit weighted graph state (where the qubits are numbered $1,2,3$ from left to right):
\begin{eqnarray} \fl W_{\phi_{12}}^{(12)} W_{\phi_{23}}^{(23)} W_{\phi_{13}}^{(13)} \Ket{+++} 
 = \frac{1}{\sqrt{8}}\big(\Ket{000} + \Ket{100} + \Ket{010} + e^{i\phi_{12}}\Ket{110} + \nonumber\\
\qquad + \Ket{001} + e^{i\phi_{13}}\Ket{101} + e^{i\phi_{23}}\Ket{011} + e^{i(\phi_{12}+\phi_{13}+\phi_{23})}\Ket{111}\big) \nonumber
\end{eqnarray}

For \textit{every} pair $a,b$ of spins, there is a phase gate with a phase $\phi_{ab}=\phi_{ba}$. The key observation and starting point of the work in \cite{spin_chain_long_range} is that even for very large $N$, we can efficiently calculate any reduced density matrix for a subset $A\subset\{1,2,\dots,N\}$ of the qubits as long as the number of qubits in $A$ (i.\,e. the number of qubits not traced over) is low. This calculation is efficient in the sense that the time requirement scales only polynomially in $N$ (though exponentially in $|A|$). This is remarkable because in the generic case, the time to calculate a reduced density matrix is exponential in $N$, and most classes of states which allow for calculation of reduced density matrices in polynomial times are bounded in the amount of entanglement that they can contain. Especially in the case of matrix product states, this fact is the dominant reason why DMRG cannot be applied successfully for certain settings \cite{DMRG_seen_from_QIT}. Weighted graph states, on the other hand, are not bounded in the amount of their entanglement, as shall be explained in Sec.\ \ref{prop}.

There is no guarantee that these states spread through those parts of the Hilbert space which are of interest for us, and hence, we add as many further degrees of freedom to the form (\ref{simpleWGS}) as possible without losing the ability to efficiently calculate reduced density matrices. As will be demonstrated in Sec.\ \ref{calc_rdm}, the following additions do not hinder the efficiency of the reduced density matrix evaluation: (i) Let the phase gates act not simply on $\Ket{+}^{\otimes N}$, but on any $N$-qubit product state. (ii) Even weighted superpositions of $m$ product states can be treated, provided $m$ is small. (iii) After the phase gates, arbitrary local unitaries may be applied.

\subsection{Parametrisation}

Deviating from the treatment in \cite{WGSMinPRL}, we develop the formulae not just for spin-1/2 particles but, more generally, for $n$-level systems, i.\,e. our states live in a Hilbert space $\mathcal{H}=(\mathbb{C}^n)^{\otimes N}$. 

\subsubsection{Superposition of product states}

We start with a superposition of $m$ product states, which we write
\begin{eqnarray} \on{nrm}\sum_{j=1}^m \bigotimes_{a=1}^N \alpha_j \left(\Ket{0} + d_{a,1}^j\Ket{1} + d_{a,2}^j\Ket{2} + \dots + d_{a,n-1}^j\Ket{n-1}\right) \nonumber \\ 
\qquad = \on{nrm}\sum_{j=1}^m\alpha_j \bigotimes_{a=1}^N \sum_{s\in\mathbb{S}} d_{as}^j\Ket{s}. \label{nlevelprodstate} \end{eqnarray}
The operator $\on{nrm}$ denotes normalisation: $\on{nrm} \Ket{\psi} := \Ket{\psi} / \| \Ket{\psi}\|$. To facilitate notation, we also introduced
\begin{eqnarray}
V &:= \{1, 2, \dots, N\}\qquad&\text{(set of spins)} \nonumber \\
\mathbb{S} &:= \{0, 1, \dots, n-1\}\qquad&\text{(set of levels)} \nonumber
\end{eqnarray}
As we normalise afterwards, we can fix the coefficient in front of $\Ket{0}$ to 1: $d_{a,0}^j\equiv 1$ for all $a,j$. It will also be useful later to introduce the \textit{deformation operators}
\[ D_{\mathbf{d}} := \sum_{s\in\mathbb{S}} d_s\Ket{s}\Bra{s},\qquad\text{with } \mathbf{d}=(d_0,d_1,\dots,d_{n-1}) \]
and the $n$-level $\Ket{\boldsymbol{+}}$ state
\[ \Ket{\boldsymbol{+}} := \frac{1}{\sqrt{n}}\sum_{s\in\mathbb{S}} \Ket{s}, \]
such that the state (\ref{nlevelprodstate}) can now be written in the forms
\bEq \on{nrm} \left( \bigotimes_{s\in\mathbb{S}} D_{\mathbf{d}_a^j}\right)\Ket{\boldsymbol{+}}^{\otimes N} =
\on{nrm} 
\sum_{\mathbf{s}\in\mathbb{S}^N} 
\left(\prod_{c\in V} d_{c,s_c}^j\right) \Ket{\mathbf{s}} \label{prod_d}
\eEq

\subsubsection{Phase gate}

We entangle these product states by applying onto each pair $a, b$ of spins a generalisation $W_\Phi$ of the 2-level phase gate $W_\phi$ from Eq.\ (\ref{phase_gate_1_2}). We want to define $W_\Phi$ as general as possible, but have to meet three constraints: (i) All $W_\Phi$ have to commute (because otherwise the calculation of reduced density matrices explained later in Sec.\ \ref{calc_rdm} does not work). Hence, they have to be diagonal. (ii) $W_\Phi$ has to be unitary (for the same reason). Hence, the entries in its diagonal have to be pure phases. (iii) $W_\Phi$ should not have any parameters which can be absorbed without loss of generality into the $d_{as}^j$. To see, which these are, let us look at the example of $n=3$: 
\[\fl W_\Phi \left[
\left( \begin{array}{c}\beta_0\\ \beta_1\\ \beta_2\end{array} \right) \otimes
\left( \begin{array}{c}\gamma_0\\ \gamma_1\\ \gamma_2\end{array} \right)
\right] =
\left( \begin{array}{c}\Phi^{00}\, \beta_0\gamma_0\\ \Phi^{01}\, \beta_0\gamma_1\\ \Phi^{02}\, \beta_0\gamma_2\\ \Phi^{10}\, \beta_1\gamma_0\\ \Phi^{11}\, \beta_1\gamma_1\\ \Phi^{12}\, \beta_1\gamma_2\\ \Phi^{20}\, \beta_2\gamma_0\\ \Phi^{21}\, \beta_2\gamma_1\\ \Phi^{22}\, \beta_2\gamma_2\\ \end{array} \right) 
= 
\left( \begin{array}{c}\zeta_{01}\\ \zeta_{02}\\ \zeta_{03}\\ \zeta_{10}\\ \zeta_{11}\\ \zeta_{12}\\ \zeta_{20}\\ \zeta_{21}\\ \zeta_{22}\\ \end{array} \right) 
\]
If one is given the $\zeta_{st}$ and can choose the $\beta_s$ and $\gamma_t$ at will, one does not need the freedom to set all entries in $W_\Phi$. It suffices to have 4 phases:
\bEq W_\Phi = \on{diag}\, (1, 1, 1,\, 1, \rme^{\rmi\Phi^{11}}, \rme^{\rmi\Phi^{12}}, \, 1, \rme^{\rmi\Phi^{21}}, \rme^{\rmi\Phi^{22}}).\label{WPhi3}\eEq
In general, for $n$ levels, one needs to specify $(n-1)^2$ phases for each phase gate $W_\Phi$. We denote the phases by a $(n-1)\times(n-1)$ matrix $\Phi_{ab}$ (with elements $\Phi_{ab}^{st}$) and have
\[W_\Phi = \mathbb{1}_{n\times n} \oplus \bigoplus_{s=1}^{n-1} \on{diag} 
\left(1, \rme^{\rmi\Phi^{s1}}, \rme^{\rmi\Phi^{s2}}, \dots, \rme^{\rmi\Phi^{s,n-1}}\right). \]
Defining $\Phi^{s0}\equiv 0$ and $\Phi^{0t}\equiv 0$ for all $s,t\in\mathbb{S}$, we can simply write
\bEq W_\Phi = \sum_{s,t\in\mathbb{S}} \rme^{\rmi\Phi^{st}} \Ket{st}\Bra{st} \label{WPhi} \eEq

Our variational states now take the following form:
\bEq \Ket{\Psi(\mathbf{x})} = \on{nrm} \left(\bigotimes_{a=1}^N U_a\right) \sum_{j=1}^m \alpha_j \left(\prod_{\substack{a,b\in V \cr a<b}}W_{\Phi_{ab}}^{(ab)}\right) \bigotimes_{a=1}^N \sum_{s\in\mathbb{S}} d_{as}^j\Ket{s}. \label{VarStateFull} \eEq
The vector $\mathbf{x}$ is a concatenation of all the parameters that are present in the right-hand side, i.\,e. the (real) parameters of $\mathbf{x}$ contain the real and imaginary parts of the complex scalars $d_{as}^j$ and $\alpha_j$, the (real) entries $\Phi_{ab}^{st}$ of the phase matrices, and the parameters describing the $N$ local unitaries $U_a\in SU(n)$, $a=1,\dots,N$.

\subsubsection{Parametrisation of the unitaries}

Next, we need to choose a parametrisation of $SU(n)$ in order to describe the unitary matrices $U$. For this, we use an isomorphism between the set $SU(n)$ of unitary $n\times n$ matrices $U$ and the set of Hermitian $n\times n$ matrices $A$ because Hermitian matrices are easy to parametrise. We could use (a) the matrix exponentiation $U=\exp{\rmi A}$ or (b) the \textit{Cayley transform} (introduced 1846 by Cayley, see e.\,g. \cite{CayleyTransform})
\bEq U=(\rmi\mathbb{1}+A)(\rmi\mathbb{1}-A)^{-1}.\label{Cayley_trafo} \eEq
To calculate these expressions numerically, we need, for (a), a matrix diagonalisation and, for the matrix invertion in (b), an LU factorisation \cite{NumLinAlg}. We choose the Cayley transform, not only because it is slightly faster, but especially because we will later have to evaluate the derivatives of $U$ with respect to its parameters, and while this is very involved for (a) \cite{deriv_matr_exp}, it is rather trivial for (b) \cite{deriv_inv_matr}. (A disadvantage seems to be on the first glance that the Cayley transform is undefined if $A$ has -1 as eigenvalue, because then, $(\rmi\mathbb{1}-A)$ cannot be inverted. The algorithm will not, however, converge to this case, and if it happened to hit on it, the program would abort.)

\subsubsection{Parameter count}

Let us now count the number $K$ of real parameters needed to describe a state $\Ket{\Psi(\mathbf{x})}$:
\begin{compactitem}
\item For each phase gate, we need $(n-1)^2$ real numbers. In case of one phase matrix for each pair of spins, there are $N(N-1)/2$ gates.
\item For the deformations, i.\,e., the specification of the initial product states, we need $2mN(n-1)$ real numbers.
\item For the superposition coefficients, $2m$ reals.
\item An $n\times n$ Hermitian matrix is specified by $n(n-1)/2$ complex entries in one of the triangles above or below the diagonal and $n$ real entries in the diagonal. Hence, we need for the $N$ unitaries a total of $Nn(n+1)/2$ real parameters.
\end{compactitem}

Thus, the number of parameters is
\begin{eqnarray} K &=  (n-1)^2 \frac{N(N-1)}{2} + 2mN(n-1) + 2m + N\frac{n(n+1)}{2} \nonumber \\
&= O(N^2n^2 + Nnm). \label{num_param_unsymm}
\end{eqnarray}

\subsection{Entanglement properties}\label{prop}

As already mentioned an important motivation for this work was the goal to find a class of states which exhibit strong entanglement over arbitrary distances that is somewhat ``generic''. After all, the limited ability to describe such entanglement is a common shortcoming of many approximation methods for many-body quantum mechanics. For the case of DMRG, this has been studied in detail in Ref. \cite{DMRG_seen_from_QIT}. There, it was shown that the matrix product states that arise during DMRG can be understood as ``projections'' from an auxilliary linear quantum system of the valence bond solid type \cite{vbs_for_qc}. Hence, whenever one cuts the matrix product states ``chain'' into two parts, the blockwise entanglement (i.\,e., the entropy of the reduced density matrix of one of either part) is bounded by $2 \log_2 D$, where $D$ is the dimension of the auxilliary spins, which is equal to the number of ``kept states'' in DMRG parlance or the matrix size in the matrix product state picture. This explains why DMRG performs not too well when applied to long 1D systems with long-range entanglement or, more precisely, to systems where the blockwise entanglement grows with the block size.

A scaling of the entanglement is hardly avoidable when treating systems with more than one dimension. According to the various ``area law'' theorems and conjectures, for most systems the entanglement of a block versus the rest of the system scales linearly with the area of the interface between this block and the rest \cite{AreaLaw1,AreaLaw2,AreaLaw_Fermions,area_law_harmonic_lattice, holographic_principle_in_spin_lattices}. Hence, for, say, a 2D system, the entanglement scales linearily with the surface area of the block and matrix product states are unable to render this feature without their matrix size growing quite fast. There are ways of replacing the matrices with higher-rank tensors to keep up with the area law, yielding so-called projected entangled pair states (PEPSs) \cite{Verstr_Renorm} but the formalism of these is rather tedious and grows more complicated with increasing spatial dimension. Also, PEPSs cannot go beyond the area law and are hence still unable to treat systems that do not follow the area law, i.e., show entanglement that scales superlinearly with the block surface, which typically is the case in critical and certain disordered systems \cite{entanglement_crit_1D_gapless,entanglement_crit_1D,area_law_and_statistics,area_law_PEPS,entanglement_scaling_time_evol,entanglement-crit-2d}.

We hope that our variational method turns out to be a viable complementary method especially to this ``PEPS'' generalization of DMRG. To see how this claim may be substantiated, note that in the description of our states, the geometry of the system has not entered yet. Every spin is connected to every other spin by a phase gate, and we can thus modell any geometry, i.\,e., any scheme of neighboring relations.
The entanglement of a block of $M$ spins w.\,r.\,t.\ the rest of the system (with $N-M$ spins) can scale with the number of spins $M$, i.e. with the {\em volume} and not with the surface area of the block \cite{SpinGasPRL}. Thus, the blockwise entanglement can reach the maximum value that is possible in the given Hilbert space. Other entanglement measures such as localizable entanglement between pairs of spins and also two-point correlation functions can reach their maximum value (independent of the distance), but can also show exponential or polynomial decay \cite{spin_chain_long_range}. This is already evident from the fact that 2D cluster states are within our variational class, and they
reach maximum entanglement in several senses \cite{UniversalResourcePRL}, e.\,g.\ the localizable entanglement between all pairs of spins is one.

\subsection{Making use of symmetries} \label{symms}
\nopagebreak
\subsubsection{Symmetrising the phases} \label{phasesymms}
\nopagebreak
The quadratic scaling of $K$ with the number $N$ of spins (lattice sites) in \req{num_param_unsymm} can be reduced to a linear scaling in case of a system Hamiltonian with translational symmetry. This is because in this case it is reasonable to assume that we do not lose precision if we let the phase matrices depend not on the absolute positions of the spins $a$ and $b$ but only on the position of $b$ relative to $a$. More precisely, we introduce a mapping $\nu: V\times V\rightarrow \{1,\dots,R\}$, that gives the \textit{phase index} for the spin pair $a, b$: the phase gate that is applied on the pair $(a,b)$ shall be the phase matrix with number $\nu(a,b)$, and $R$ is the total number of phase matrices. The 4th-order tensor $\Phi^{r_1r_2}_{ab}$ now becomes a 3rd-order tensor $\Phi^{r_1r_2}_{\nu(a,b)}$. 

\begin{figure}
\includegraphics[width=.4\textwidth]{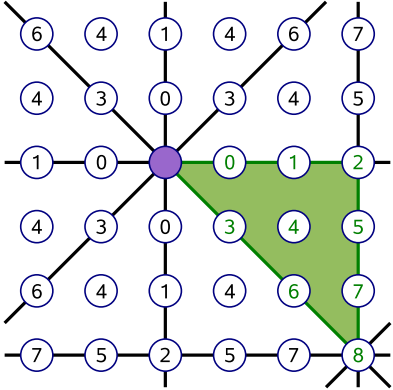}
\caption{For a system on an $L\times L$ square lattice with periodic boundary conditions and a system Hamiltonian that is invariant under the lattice's symmetry group, only $R=O(N)$ phase matrices are needed. The numbers indicate the numbering of these matrices with phase indices $\nu=0,\dots,R-1$. The circles denote sites of a $6\times 6$ lattice. In constructing a variational state (\ref{VarStateFull}) on this lattice, phase gates $W_{\Phi_\nu}$ are performed on any pair of sites, where the phase gate acting on the purple shaded site and a site marked with the number $\nu$ uses the phase matrix $\Phi_\nu$. Translation of these markings show the phase indices for other site pairs. Note, how due to the rotation and reflection symmetries of the square lattice, the pattern of phase indices is repeated eight times.}
\label{phase_indices_2D_square}
\end{figure}

The mapping $\nu$ has to be constructed such that two pairs of spins, $(a,b)$ and $(c,e)$, get the same index, $\nu(a,b)=\nu(c,e)$, if and only if the pair $(a,b)$ can be mapped onto $(c,e)$ by a symmetry transformation that leaves the system Hamiltonian invariant. For the common case of a Hamiltonian that is a sum of identical terms which each act on one bond (i.\,e., connection of lattice sites), this is the symmetry group of the lattice. In the case of a square lattice with $N=L\times L$ sites on periodic boundary conditions (PBC), only\footnote{The brackets $\lfloor\cdot\rfloor$ denote the floor function.} 
\[R=\frac{1}{2}\left\lfloor\frac{L}{2}\right\rfloor \left(\left\lfloor\frac{L}{2}\right\rfloor-1\right) = O(N)\]
phase matrices are needed as can be seen from Fig.\ \ref{phase_indices_2D_square}, and thus, we need only $K=O(N)$ parameters.

Note also, that $\nu$ is naturally symmetric, $\nu(a,b)=\nu(b,a)$, and that this has to be reflected by a like symmetry of $\Phi$ \wrt its \textit{upper} indices: $\Phi_\nu^{r_1r_2} = \Phi_\nu^{r_2r_1}$, which must be imposed explicitely.

\subsubsection{Full symmetrisation} \label{fullsymm}

For a symmetric Hamiltonian, it seems natural to reflect this symmetry not only in the phases $\Phi$, but also in the local, site-dependent properties, i.\,e., in the local unitaries $U_a$ and the deformation parameters $\mathbf{d}_a^j$. In case of full translation symmetry, one may want to completely drop the dependence of these on the site index $a$. This does indeed reduce the number $K$ of parameters significantly, but not as dramatically as in the case of phase symmetrisation. The latter reduced the scaling of $K$ from $\Or(N^2)$ to $\Or(N)$, while further symmetrisation of the other parameters cannot change $K=\Or(N)$. On the other hand, the time required to calculate the energy of a given state is reduced by a factor $\Or(N)$ in the fully symmetric case, as one needs to evaluate it for only one elementary cell of the lattice.

A good reason not to impose full symmetrisation nevertheless is the observation that for many systems, the ground state does not necessarily obey the full symmetry of the Hamiltonian due to spontaneous symmetry breaking. Even though in such a case, the ground state must be degenerate, and at least one state within the ground subspace must obey the full symmetry, this state is unlikely to be the state that is easiest to approximate within the chosen class of variational states. To give an example: The ground state of the antiferromagnetic Ising chain without transverse field is $\alpha\Ket{0101\dots} + \beta\Ket{1010\dots}$, for any $\alpha,\beta$ with $|\alpha|^2+|\beta|^2=1$. Only for $\alpha=\beta$ the state is invariant under a translation of one site. However, the state most easily approximated is $\alpha=1,\beta=0$ (or vice versa), as it is a product state, while any other state contains long-range entanglement. If we imposed full translational symmetry onto the states, our algorithm would likely fail to find 
a good state. However, the example suggests a compromise between flexibility and low number of parameters: We make the local properties $U_a$ and $\mathbf{d}_a^j$ periodic in a way that matches the expected periodicity of the spontaneously-broken ground state, e.\,g., in the case of the Ising chain, we may use one common unitary and one common deformation vector for all odd sites, and another unitary and another deformation vector for all even sites. However, our numerical experiments showed that this does not work particularly well: the enforcement of such symmetries introduces very many additional local minima which trap the minimzation routine much too soon. The intuitive reason for this is that enforcing the symmetrie amount to a cut through the energy landscape of the parameter space which seems to divide meandering troughs into seperated basins.

Let us nevertheless mention two more possibilities to even further reduce the parameter scaling. (i) We can make the phase index mapping $\nu$ such that it does not depend on the geometric relation as in Fig.\ \ref{phase_indices_2D_square} but just on the scalar number of lattice steps that separates the spins, the number of phase indices scales linearly only with the length $L$, not with the number of sites $N=\Or(L^\mathcal{D})$ (where $\mathcal{D}$ is the dimension of the system). Together with a full or periodic symmetrization of the local properties, we reach a scaling of the number of parameters $K=\Or(L)$, which allows for a quick treatment even of 3D systems of moderate size. The accuracy achieved this way is, however, very modest.

(ii) Often, one may expect long-range entanglement to be supressed exponentially. Then one can choose a distance threshold and fix to zero all phases betweens spins with a distance above this threshold. The threshold will typically be chosen of the order of the entanglement length, and as the latter usually does not increase strongly with the system size (except at criticality) one can save considerably on the number of parameters.

\section{Evaluating observables} \label{calc_rdm}

In order to evaluate an observable $\mathcal{O}$ with support on $A\subset V$, we need to evaluate
\[\langle\mathcal{O}\rangle_\mathbf{x} = \tr \mathcal{O} \rho_A\]
with
\[ \rho_A := \tr_{V\backslash A} \Ket{\Psi(\mathbf{x})}\Bra{\Psi(\mathbf{x})}. \]
As we shall see now, $\rho_A$ can be calculated in time polynomial in the number $N-|A|$ of spins, the number $n$ of levels per spin and the number $m$ of superpositions, but exponential in the number $|A|$ of spins not traced over. Hence, the expectation value $\langle\mathcal{O}\rangle_\mathbf{x}$ of observables can be calculated efficiently as long as $\mathcal{O}$ is a sum of terms with small support.

In particular, we need this algorithm to evaluate the energy $\Bra{\Psi(\mathbf{x})}H\Ket{\Psi(\mathbf{x})}$, as this is the quantity we wish to minimise. Thus, due to the scaling properties just mentioned, we require that the system Hamiltonian can be written as sum of terms with small support (as it is the case nearly always).

\subsection{A pair of spins} \label{deriv_rho_ab}

To keep notation simple, we only derive the procedure to obtain the two-spin density matrix ($A=\{a,b\}$)
\bEq \rho_{ab} := \rho_{\{a,b\}} = \tr_{V\backslash\{a,b\}} \Ket{\Psi(\mathbf{x})} \Bra{\Psi(\mathbf{x})}.\label{rho_ab_start} \eEq
This is a generalisation of the work done in \cite{spin_chain_long_range} for spin-1/2. A further generalisation to more than two spins is easy and its result will be given at the end of this section.

The spins that we do not trace over are denoted $a$ and $b$. We start by inserting \req{VarStateFull} into \req{rho_ab_start} and pull as much as possible out of the partial trace:
\begin{eqnarray}
\rho_{ab} = \on{nrm} \left(U_a\otimes U_b\right) W_{\Phi_{ab}}\times \nonumber\\
\qquad {}\times \left(\sum_{j,k=1}^m \alpha_j \alpha_k^* 
\left( D_{\mathbf{d}_a^j}\otimes D_{\mathbf{d}_b^j}\right) 
\rho_{ab}^{jk}
\left( D_{\mathbf{d}_a^k}\otimes D_{\mathbf{d}_b^k}\right)^\dagger\right) \times \label{rho_ab_S} \\
\qquad\qquad {}\times W_{\Phi_{ab}}^\dagger \left(U_a\otimes U_b\right)^\dagger \nonumber.
\end{eqnarray}
Here, the operator $\on{nrm}$ again means normalisation, now defined as $\on{nrm}\rho := \rho / \tr\rho$, and the inner term $\rho_{ab}^{jk}$ contains anything that cannot be pulled out of the partial trace:
\bEq \rho_{ab}^{jk} = \tr_{V\backslash\{a,b\}} \Ket{\psi_{ab}^j} \Bra{\psi_{ab}^k\vphantom{\psi_{ab}^j}} \label{rhoabjk_def} \eEq
with
\[
\Ket{\psi_{ab}^j} = 
\left(\prod_{c\in V\backslash\{a,b\}} W_{\Phi_{ac}}^{(ac)} W_{\Phi_{bc}}^{(bc)} \right)  \bigotimes_{c\in V\backslash\{a,b\}} \,\, \sum_{s\in\mathbb{S}} d_{c,s}^j \Ket{s}, \\
\]
which is, due to Eqs.~(\ref{prod_d}) and (\ref{WPhi}),
\[ \Ket{\psi_{ab}^j} =
\sum_{\mathbf{s}\in\mathbb{S}^N} 
\left( \prod_{c\in V\backslash\{a,b\}} \rme^{\rmi\Phi_{ac}^{s_a s_c}} \rme^{\rmi\Phi_{bc}^{s_b s_c}} d^j_{s,s_c} \right)
\Ket{\mathbf{s}}.
\]
Note that in the trace (\ref{rhoabjk_def}) all the phase gates $W_{\Phi_{ce}}$ with $c,e\notin\{a,b\}$ cancel with their Hermitian conjugate, as do all the local unitaries $U_c$, $c\notin\{a,b\}$. Hence, $\rho_{ab}$ depends only on a subset of the parameters.

In order to take the trace in \req{rhoabjk_def}, we have to sum over all states $\Ket{\underline{\mathbf{q}}}$, $\underline{\mathbf{q}}\in\mathbb{S}^{N-2}$, where the underline denotes that the components of $\underline{\mathbf{q}}$ are \textit{not} indexed $(s_1,s_2,\dots,s_{N-2})$ but rather using the elements of $V\backslash \{a,b\}$ as indices. We get
\begin{eqnarray}
\fl \rho_{ab}^{jk} 
&= \sum_{\underline{\mathbf{q}}\in\mathbb{S}^{N-2}} 
\Braket{\underline{\mathbf{q}} | \psi_{ab}^j} \Braket{\psi_{ab}^k \vphantom{\psi_{ab}^j}| \underline{\mathbf{q}}} \nonumber \\
\fl &= \sum_{\underline{\mathbf{q}}\in\mathbb{S}^{N-2}} \sum_{\mathbf{s},\mathbf{s}'\in\mathbb{S}^N}
\Braket{\underline{\mathbf{q}} | \mathbf{s}} \Braket{\mathbf{s}' | \underline{\mathbf{q}}} \times \nonumber \\
\fl & \qquad\qquad \times \prod_{c\in V\backslash\{a,b\}} d_{cs_c}^j {d_{cs_c}^k\!\!}^*
\exp \left[\rmi \left(\Phi_{a\vphantom{b}c}^{s_a s_c} + \Phi_{bc}^{s_b s_c} - \Phi_{a\vphantom{b}c}^{s_a' s_c'} - \Phi_{bc}^{s_b' s_c'}\right)\right] \nonumber \\
\fl &= \sum_{\mathbf{r},\mathbf{r}'\in\mathbb{S}^2} \Ket{\mathbf{r}}\Bra{\mathbf{r}'}
\sum_{\underline{\mathbf{q}}\in\mathbb{S}^{N-2}} \prod_{c\in V\backslash \{a,b\}} d_{cs_c}^j {d_{cs_c}^k\!\!}^*
\exp \left[\rmi \left(\Phi_{a\vphantom{b}c}^{r_1 q_c} + \Phi_{bc}^{r_2 q_c} - \Phi_{a\vphantom{b}c}^{r_1' q_c} - \Phi_{bc}^{r_2' q_c}\right)\right]
\nonumber
\end{eqnarray}
In the last line, we can exchange sum and product in the following manner without changing the expression:
\[ \sum_{\underline{\mathbf{q}}\in\mathbb{S}^{N-2}} \,\, \prod_{c\in V\backslash \{a,b\}} \longrightarrow
\prod_{c\in V\backslash \{a,b\}} \,\, \sum_{q_c\in\mathbb{S}}. \]
This gives
\bEq \fl \rho_{ab}^{jk} = \sum_{\mathbf{r},\mathbf{r}'\in\mathbb{S}^2} \Ket{\mathbf{r}}\Bra{\mathbf{r}'}
\prod_{c\in V\backslash \{a,b\}}  \underbrace{ \sum_{q\in\mathbb{S}} d_{cq}^j {d_{cq}^k\!\!}^*
\exp \left[\rmi \left(\Phi_{a\vphantom{b}c}^{r_1 q} + \Phi_{bc}^{r_2 q} - \Phi_{a\vphantom{b}c}^{r_1' q} - \Phi_{bc}^{r_2' q}\right)\right]}_{:= \left(\rho_{ab,c}^{jk}\right)_{\mathbf{r},\mathbf{r}'}}\label{rho_ab_jk_pre} \eEq

The sum over $\mathbb{S}$ has $n$ terms, and $N-2$ such sums are multiplied. Hence, in order to calculate one matrix element of $\rho_{ab}^{jk}$ we have to evaluate the underbraced term $(N-2)n$ times. This is the origin of the promised polynomial scaling for the calculation of expectation values.

Recall that $\rho_{ab}^{jk}$ is an $n^2\times n^2$ matrix. We will make this more explicit by writing the product as Hadamard product. The Hadamard product, denoted $\odot$, is defined as the component-wise multiplication of matrices, $(B\odot C)_{ij} := B_{ij} C_{ij}$. Its identity, denoted $\mathbb{1}_\odot$, is the matrix having 1 as all of its elements. Using this, we can rewrite the previous equation in a very compact form:
\bEq \rho_{ab}^{jk} = \bigodot_{c\in V\backslash \{a,b\}} \rho_{ab,c}^{jk} \label{rho_compact} \eEq
where the matrix elements of $\rho_{ab,c}^{jk}$ are given by the underbraced term in \req{rho_ab_jk_pre}.
Each factor of the Hadamard product can be understood as resulting from the interaction of the spins $a$ and $b$ with one spin $c$ from $V\backslash\{a,b\}$. These factors can be calculated seperately because the interaction between two different spins in $V\backslash\{a,b\}$ may be and is ignored due to the cancellation of all phase gates within $V\backslash\{a,b\}$. (Cf. the remark after Eq.\ (\ref{rhoabjk_def}).)

To make this more concrete, let us look at the simple case of $n=2$. Then (Recall that $d_{a,0}^j\equiv 1$, and $\Phi^{s,t}=0$ for $s=0$ or $t=0$.)
\[ \rho_{ab}^{jk} = \bigodot_{c\in V\backslash\{a,b\}} \left( \mathbb{1}_\odot + d_c^j {d_c^k}^* \tilde\rho_{ab,c} \right) \]
with
\bEq \fl \tilde\rho_{ab,c} = 
\left( \begin{array}{cccc}
1 & e^{-i\Phi_{ac}^{1,1}} & e^{-i\Phi_{bc}^{1,1}} & e^{-i(\Phi_{ac}^{1,1}+\Phi_{bc}^{1,1})} \\
\rme^{\rmi\Phi_{ac}^{1,1}} & 1 & \rme^{\rmi(\Phi_{ac}^{1,1}-\Phi_{bc}^{1,1})} & e^{-i\Phi_{bc}^{1,1}} \\
\rme^{\rmi\Phi_{bc}^{1,1}} & \rme^{\rmi(-\Phi_{ac}^{1,1}+\Phi_{bc}^{1,1})} & 1 & e^{-i\Phi_{ac}^{1,1}} \\
\rme^{\rmi(\Phi_{ac}^{1,1}+\Phi_{bc}^{1,1})} & \rme^{\rmi\Phi_{bc}^{1,1}} & \rme^{\rmi\Phi_{ac}^{1,1}} & 1 
\end{array}
\right), \label{rhoabc}\eEq
which is the formula given in \cite{spin_chain_long_range}.

\subsection{Several spins}

For reference, we give the result for density matrices for not simply two spins $a,b$, but arbitrary numbers of spins, given in a set $A$:
\begin{eqnarray}
\fl \rho_A &:= \tr_{V\backslash A} \Ket{\Psi(\mathbf{x})}\Bra{\Psi(\mathbf{x})}\nonumber\\
\fl &= \on{nrm} \left( \bigotimes_{a\in A} U_a\right)
\left( \prod_{\substack{a,b\in A\cr a<b}} W_{\Phi_{ab}}^{(\tau(a),\tau(b))}\right) \times\nonumber\\
\fl&\qquad\times\left[\sum_{j,k=1}^m \alpha_j \alpha_k^* \left( \bigotimes_{a\in A} D_{\mathbf{d}_a^j} \right)
\left(\bigodot_{c\in V\backslash A} \rho_{A,c}^{jk} \right)
\left( \bigotimes_{a\in A} D_{\mathbf{d}_a^k} \right)^\dagger\right] \times\nonumber\\
\fl&\qquad\hfill\times\left( \prod_{\substack{a,b\in A\cr a<b}} W_{\Phi_{ab}}^{(\tau(a),\tau(b))} \right)^\dagger
\left( \bigotimes_{a\in A} U_a\right)^\dagger \label{rho_A}
\end{eqnarray}
with
\bEq \rho_{A,c}^{jk} = \left( \sum_{s\in\mathbb{S}} d_{c,s}^j {d_{c,s}^k\!\!}^*\, \exp \left[\rmi \sum_{a\in A} \left(\Phi_{ac}^{r_{\tau(a)},s} - \Phi_{ac}^{r'_{\tau(a)},s}\right)\right] \right)_{\mathbf{r}, \mathbf{r}'\in\mathbb{S}^{|A|}}.\label{rho_c_jk_S}\eEq
The mapping $\tau:A\rightarrow\{1,\dots,|A|\}$ gives here the index that spin $a\in A$ gets within the density matrix $\rho_A$ (i.\,e., in the 2-spin case of $\rho_{ab}$, $\tau(a)=1$ and $\tau(b)=2$.).

It is also useful to observe that
\bEq \left( \bigotimes_{a\in A} D_{\mathbf{d}_a^j} \right)
\rho
\left( \bigotimes_{a\in A} D_{\mathbf{d}_a^k} \right)^\dagger
= \Ket{D_A^j}\Bra{D_A^k\vphantom{D_A^j}} \odot \rho \label{rho_with_Dcheck} \eEq
with
\bEq \Ket{D_A^j} = \sum_{\mathbf{r}\in \mathbb{S}^{|A|}} \Ket{\mathbf{r}}
\prod_{a\in A} d_{a,r_{\tau(a)}}^j \label{def_Ket_D_A}\eEq
This formula comes from the observation that a matrix product of a diagonal matrix, an arbitrary matrix, and another diagonal matrix can be written as Hadamard product:
\bEq \left(\on{diag} \mathbf{u}\right)\, A\, (\on{diag} \mathbf{v}) = A \odot \left(\mathbf{u} \mathbf{v}^\dagger\right) \label{diag_to_hadm_prod} \eEq

Further, for the numerics, one may use
\[ \prod_{\substack{a,b\in A\cr a<b}} W_{\Phi_{ab}}^{(\tau(a),\tau(b))} =
\on{diag} \left(\exp \left[ \rmi \sum_{\substack{a,b\in A\cr a<b}} \Phi_{ab}^{r_{\tau(a)},r_{\tau(b)}}\right]\right)_{\mathbf{r}\in\mathbb{S}^{|A|}}.
\]

\section{Demonstration for two models} \label{models}

To approximate a ground state, we have to vary the parameters in order to minimize the energy. Before we explain our techniques to achieve this we show the results of such minimizations for two different models to demonstrate the performance of our technique. The two model systems, namely the XY model and the Bose-Hubbard model, are presented in the two following subsections. For each of the two models, we have used a different implementation (see Sec.\ \ref{perf_min} for details) and different heuristics for the global minimisation. Hence, we shall use these results as examples when explaining these heuristics in Sec.\ \ref{perf_min}. As the second implementation is newer and its heuristics more sophisticated, the results for the Bose-Hubbard model are more convincing. Nevertheless, we also present our results for the XY model, as the old heuristics provides illuminating insights into important aspects of our methods behaviour. The examples with the XY model are a continuation of the examples for the Ising model (wich is a special case of the XY model) already given in \cite{WGSMinPRL}.

\subsection{The XY model with transverse field} \label{XYmodel}

The \textit{XY model with transverse field} for a system of spin-1/2 particles on a lattice is given by the Hamiltonian
\[ H = \sum_{\{a,b\}\in \mathcal{B}} \left(\frac{1+\gamma}{2} \sigma_x^{(a)} \sigma_x^{(b)} + \frac{1-\gamma}{2} \sigma_y^{(b)} \sigma_y^{(b)}\right) +
\sum_{a\in V} B \sigma_z^{(a)}, \]
where $\sigma_{x,y,z}$ are the Pauli matrices, $\mathcal{B}$ is the set of nearest neighbours, $B$ the transverse field and $\gamma$ is called the asymmetry. For $\gamma=0$, we get as special case the \textit{XX model}, and for $\gamma=1$, we get the \textit{Ising model}.

\subsubsection{One dimension (spin rings)}

\begin{figure}
\raisebox{1ex}{(a)}\includegraphics[scale=.5]{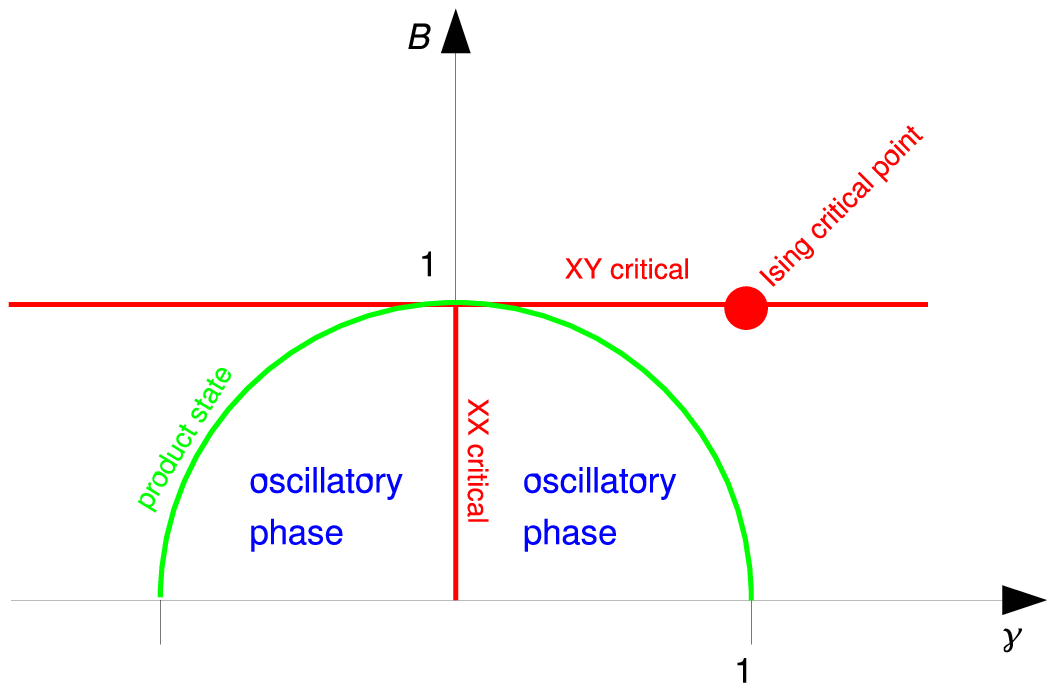}\qquad
\raisebox{1ex}{(b)}\includegraphics[scale=.5]{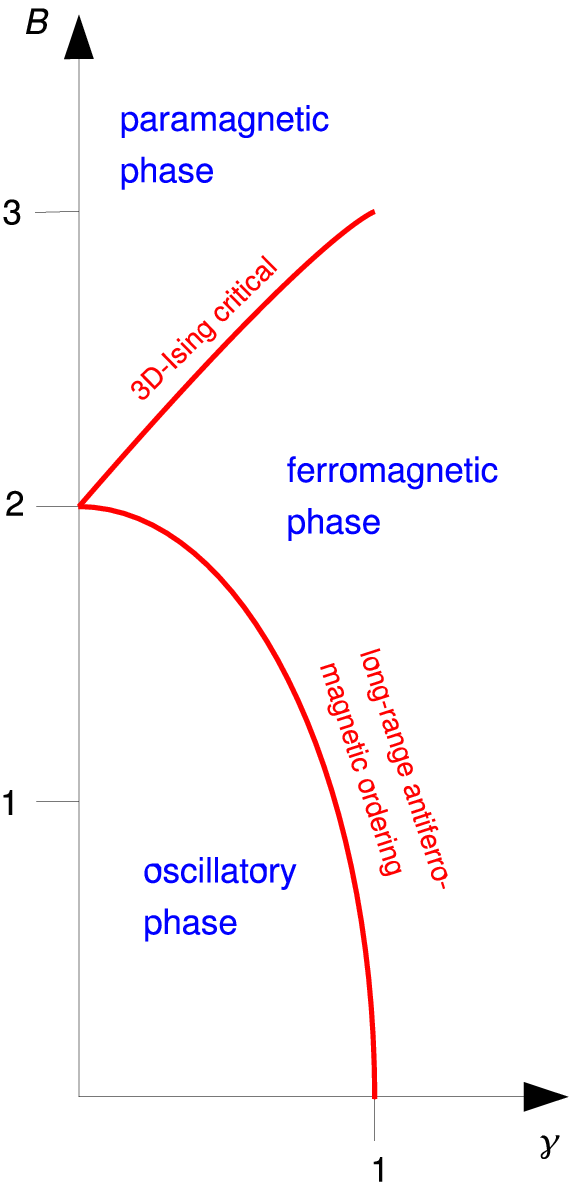}
\caption{Phase diagram of XY model, (a) for 1D (according to \cite{BarouchMcCoy}), (b) for 2D (according to \cite{XY_2D_PhaseDiagram}).}
\label{XY_Phases}
\end{figure}

For 1D, the XY model with transverse field can be diagonalised using a Jordan-Wigner and then a Bogoliubov transformation (the latter is trivial for $\gamma=1$). Correlations have been studied in early work in \cite{PfeutyIsing} (Ising) and \cite{KatsuraXYZ,BarouchMcCoy} (XY). The latter article also gives the phase diagram of the 1D XY system (reproduced in Fig.\ \ref{XY_Phases}a). The entanglement properties of these phases and their transitions have recently found much interest. The behaviour first indicated by numerical studies  \cite{Entanglement_in_Quant_Crit,EnriqueGroundSpin} was soon confirmed by analytic calculations \cite{entanglement_XX_analytical,entanglement_XY_analytical_A,entanglement_XY_analytical_B}.

\begin{figure}
\includegraphics[width=.85\textwidth]{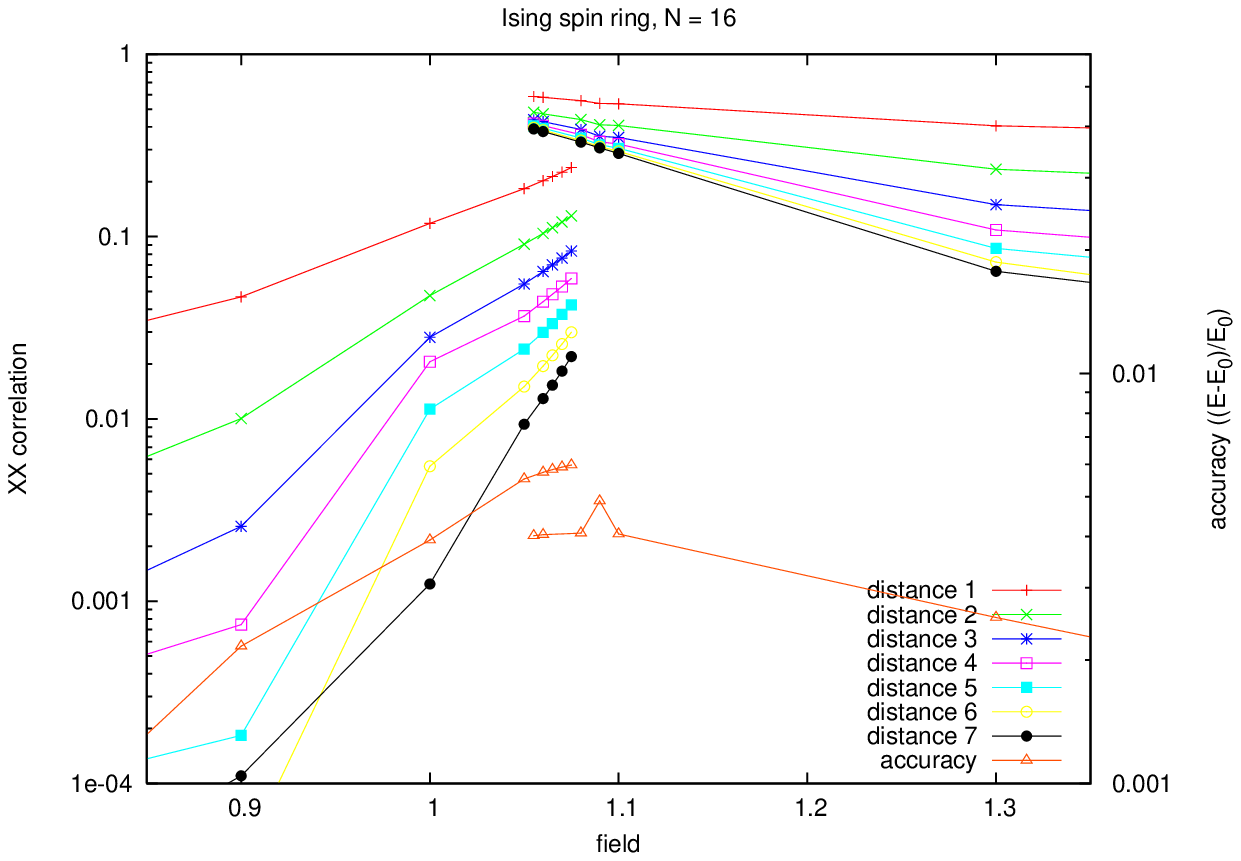}
\caption{Approximation of the ground state of an Ising ring with 16 spins, calculated for $m=4$, i.\,e., the phase gates act on superpositions of 4 product states: The orange triangles shows the deviation of the variational energy from the true ground state energy as obtained from the exact solution. The other symbols show the mean value of XX correlations for different spin-spin distances. As a guide to the eye, symbols for the same distance are conmnected by lines. It is evident that one seems to need two lines to connect the set of data points for each distance, as a single line would ``jump'' in zig-zag near the critical point. In other words: There seem to be two basins of attraction for the minimzer, corresponding to the $B<1$ and $B>1$ phase, and near the critical point, the minimizer may either fall into one basin or the other. The use of the sweeping technique, to be discussed in Sec.\ \ref{sweeping}, allows to get around this undesirable behavior and direct the minimizer to minima with smaller basins of attraction which better resemble the true ground state in the 
area of influence of the quantum phase transition. Hence, this plot is not meant to show a good result, but rather illustrate the kind of failure that motivates and nessecitates the sweeping technique.}
\label{1D_Ising}
\end{figure}

Our technique seems to be suited to study this model: the results are quite precise. Fig.\ \ref{1D_Ising} shows a transition through the Ising critical point. The curves show the XX correlations for different spin-spin distances in a ring of $N=16$ spins. 

\begin{figure}
\includegraphics[width=.95\textwidth]{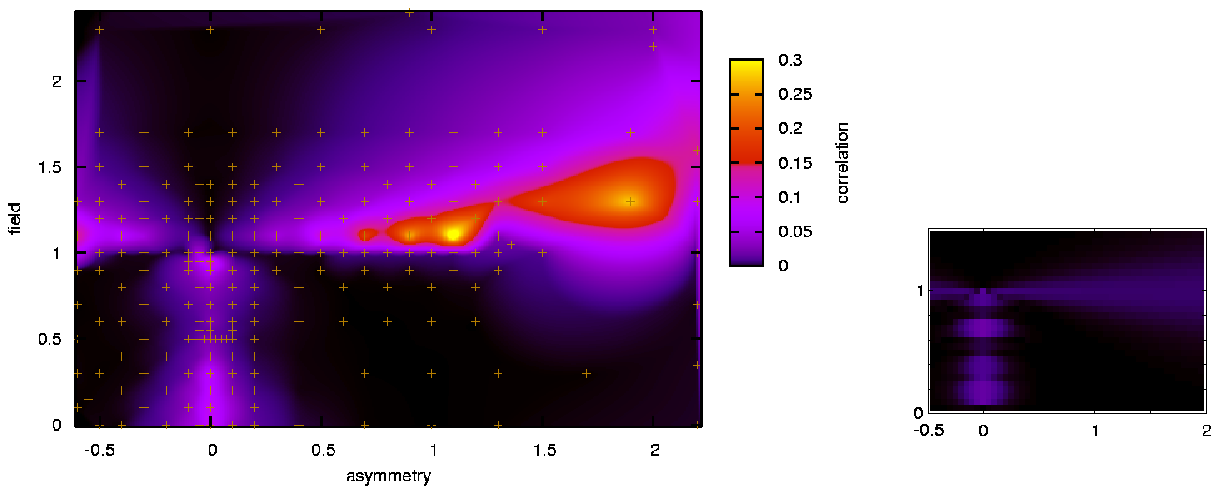}\qquad
\caption{Correlations (maximum SV of correlation matrix for distance 6) of the 1D XY model, calculated for a ring of 16 spins, and with $m=3$. The brown crosses show the positions of the data points, the colour surface in between is interpolated using Sibson's method (see Sec.\ \ref{densityplotting} and be careful to not be mislead by artefact of the interpolation, such as appareant features at regions with too sparse data points). The correlations appear at that regions that also have high entanglement in the thermodynamic limit. (Compare with Fig.\ 3 in \cite{EnriqueXY_EntropyPlot}.) However, this agreement is, unfortunately, only qualitative: A comparion with the exact result for the considered finite size case, shown in the small plot to the right, reveals that correlations are over-estimated significantly.}
\label{XY1DParamPlot}
\end{figure}

As our technique tends to spontaneously break symmetry where the true ground state does not, it makes sense to plot the two-points correlations\footnote{We either plot the XX correlations or the maximum singular value of the correlation matrix $\left(\langle \sigma_i^{(a)} \sigma_j^{(b)}\rangle - \langle \sigma_i^{(a)}\rangle \langle\sigma_j^{(b)}\rangle\right)_{i,j=x,y,z}$} for many different values of the parameters of the Hamiltonian (here: $B$ and $\gamma$) in order to spot phase transitions. We find that it works better to plot correlations for a specific distance than to estimate correlation lengths from the data because the system is still so small that the exponential decay of correlations is masked by boundary effects . Fig.\ \ref{XY1DParamPlot} shows such a plot for the 1D XY model. As is to be expected, one sees that near critical regions correlations are much stronger. (For the infinite chain, the critical regions are: XX criticality at $\gamma=0$ for $0<B<1$ and XY criticality\footnote{Strictly speaking the model is XY critical only for $B=1,\delta\neq 1$, and Ising critical for $B=\delta=1$.} for $B=1$ \cite{BarouchMcCoy}.) The spread of the areas of high correlation around the critical regions of the infinite chain looks similar areas of high entropy identified in \cite{EnriqueXY_EntropyPlot} -- compare with Fig.\ 3 in that article (and note that there, entropy is small around $\gamma=0,\;B=1$ despite the critical nature of this point -- a feature also seen in our plot of correlations.) Had we not known the critical regions, it is not merited to conclude that the system is critical where the correlations are strong, as the system size and the correlation distance is surely to small for this. We rather suggest to use a plot of this kind for a first look at a yet unstudied Hamiltonian. Regions of high correlations may suggest points in parameter space for which numerical calculations for different system sizes may give interesting results.

\begin{figure}
\qquad\includegraphics[width=.62\textwidth]{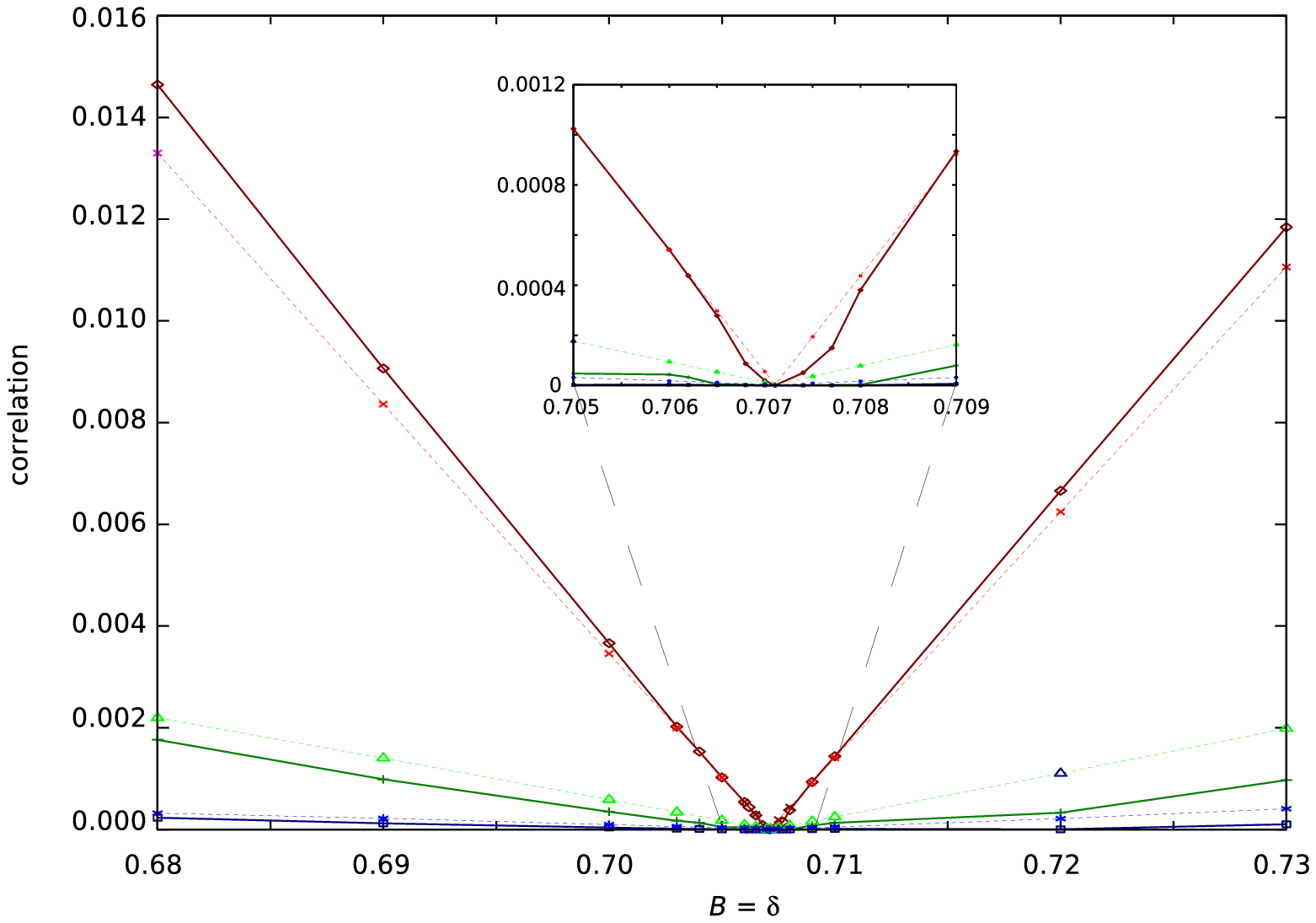}
\caption{Correlations (maximum SV of correlation matrix) in the vicinity of the Baruch-McCoy circle. Along this circle, which is defined by $B^2+\gamma^2=1$, the ground state of the XY model has product form. This is nicely reproduced by our numerics: At $B=\delta=1/\sqrt{2}$, all correlations vanish. To show this, we here plot the correlations for spin-spin distances 1 (red), 2 (green), and 3 (blue) for a cut along the line $B=\delta$, i.e., radially through the circle. The x axis is the value of $B=\delta$. Calculated for a ring of 16 spins and $m=3$ as in Fig.\ \ref{XY1DParamPlot}. The solid, dark lines are results from the variation, the dashed, light lines are exact values.}
\label{Barouch}
\end{figure}

Another interesting feature of the 1D XY model is the Baruch-McCoy circle, which is the defined by $B^2+\gamma^2=1$. On this circle, the ground state has product form \cite{BarouchMcCoy}. Our approach accurately reproduces the vanishing of all correlations as one approaches a point on the Barouch-McCoy line (Fig.\ \ref{Barouch}).

\subsubsection{Two dimensions}

The 2D XY model with transverse field has been studied in \cite{XY_2D_PhaseDiagram}. The main result of the latter treatment is illustrated by Fig.\ \ref{XY_Phases}b. 

\begin{figure}
\includegraphics[width=.73\textwidth]{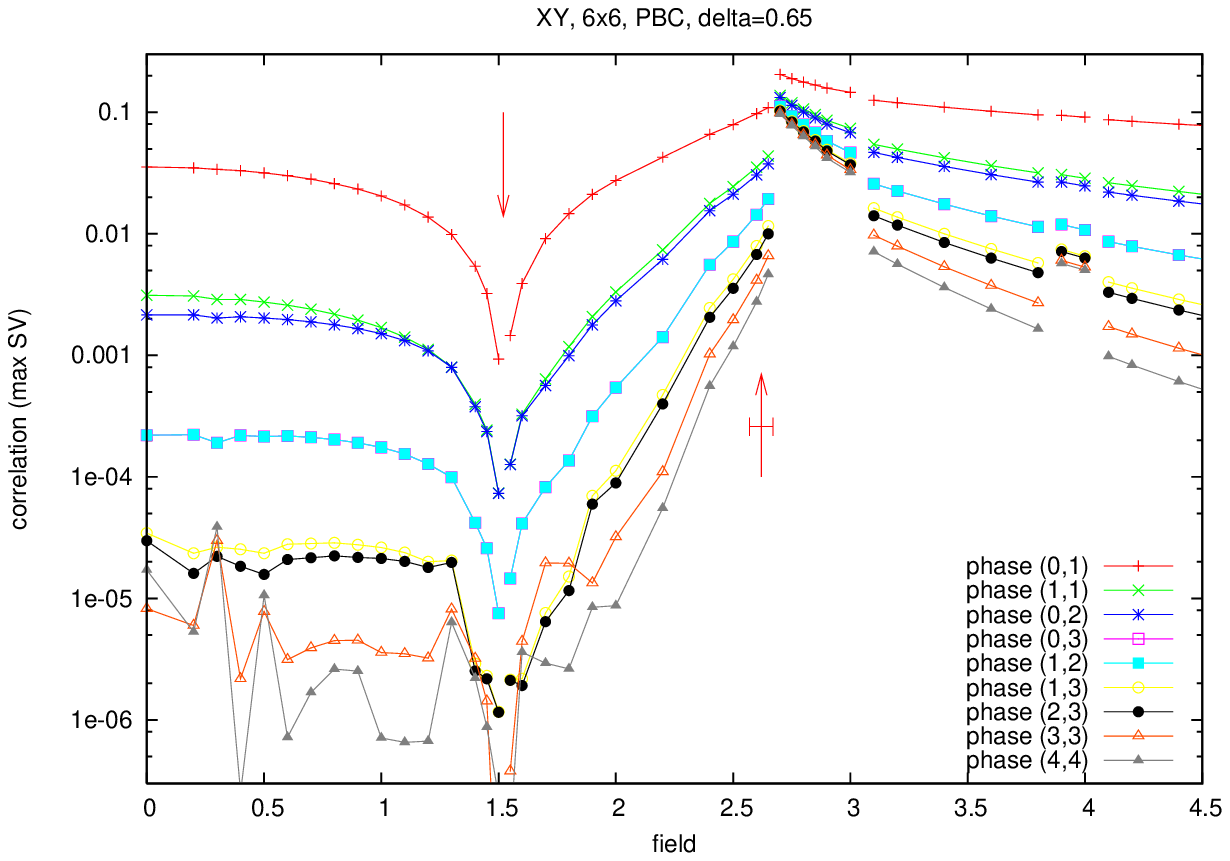}
\caption{Correlations (maximum SV of correlation matrix) for the 2D XY model, calculated for a torus of $6\times 6$ along a cross section through the phase plane of Fig.\ \ref{XY_Phases}b along the line $\gamma=0.65$ (number of superposed states: $m=3$). The red arrows show the positions of the two phase boundaries for the \textit{infinite} case (according to \cite{XY_2D_PhaseDiagram}, cf.\ Fig.\ \ref{XY_Phases}b.) As this plot has been produced without use of the sweeping technique, some instances of convergence to the wrong minimum are evident from the jumps at $B>3$. (Compare with the discussion at Fig.\ \ref{1D_Ising}). The different curves show the correlation for spin pairs with distance $(d_x,d_y)$ in $x$ and $y$ direction.}
\label{XY2D_6x6}
\end{figure}

In order to demonstrate our scheme in a 2D setting, we have done calculations for a torus (i.\,e., a square with periodic boundary conditions) of $6\times 6$ spins. We fixed the asymmetry at $\gamma=0.65$ and varied the field strength $B$ from 0 to 4.5 in order to cross both of the phase transitions indicated in Fig.\ \ref{XY_Phases}b. The results, shown in Fig.\ \ref{XY2D_6x6}, show prominent kinks at the expected positions of the phase transitions, and the correlations fall off in a roughly exponential manner with distance as expected. We still see additional jumps due to convergence into wrong basins, and this prompted us to seek a means to avoid this, namely the sweeping technique. The plots in the following section have been obtained this way and hence do not show such strong jumps.

\FloatBarrier

\subsection{Bose-Hubbard model} \label{BHmodel}

The Bose-Hubbard model is defined for a system of harmonic oscillators, arranged in a lattice, and is described by the Hamiltonian
\bEq \fl H = 
- J \sum_{\{a,b\}\in\mathcal{B}} \left( {\hat b}_{a\vphantom{b}}^\dagger {\hat b}_b^{\vphantom{\dagger}} + 
  \text{H.c.} \right)
+ U \sum_{a\in V} {\hat n}_a \left( {\hat n}_a - 1 \right) / 2 
- \mu \sum_{a\in V} {\hat n}_a 
\eEq
As before, $V$ is the set of all lattice sites, and $\mathcal{B}$ the set of all unordered pairs of nearest neighbour. The operators ${\hat b}_a^\dagger$ and ${\hat b}_a$ denote the ladder operators to create and annihilate a bosonic excitation of the oscillator at site $a$, and ${\hat n_a} = {\hat b}_a^\dagger {\hat b}_a$ is the number operator. The first term, called the \textit{hopping term} describes the ``hopping'' of an excitation from a site $a$ to a neighbouring site $b$, a process which occurs with the \textit{hopping strength} $J$. The second term describes the repulsion between several bosons on the same site. To fix our energy scale, we set the repulsion $U$ to 1 in the following, i.\,e., all dimensionless energies are to be understood in units of $U$.\footnote{When comparing with other literature, care has to be taken that many authors use the alternative convention to set $J\equiv 1$. Also, $J$ is often denoted $t$.} The last term is relevant if the particle number is not fixed, which it is in fact not in our case. Then, assigning a value to the chemical potential $\mu$ allows to choose the mean density of the ground state.

The Bose-Hubbard Hamiltonian is of interest due to its rich phase diagram, first exposed in \cite{BH_first}. While its original motivation was the description of certain structured solid state systems such as arrays of Josephson junctions, interest in the system increased significantly with the discovery that it can be realized with cold atoms in optical lattices \cite{BH_in_optical_lattices} and with the spectacular experimental demonstration of this fact \cite{SF_MI_trans_exp}, where a transition from the Mott insulator  phase to the superfluid phase and back was observed. (For a review, see \cite{Review_Optical_Lattices}).

\begin{figure}
\hspace*{-1cm}\includegraphics[width=\textwidth]{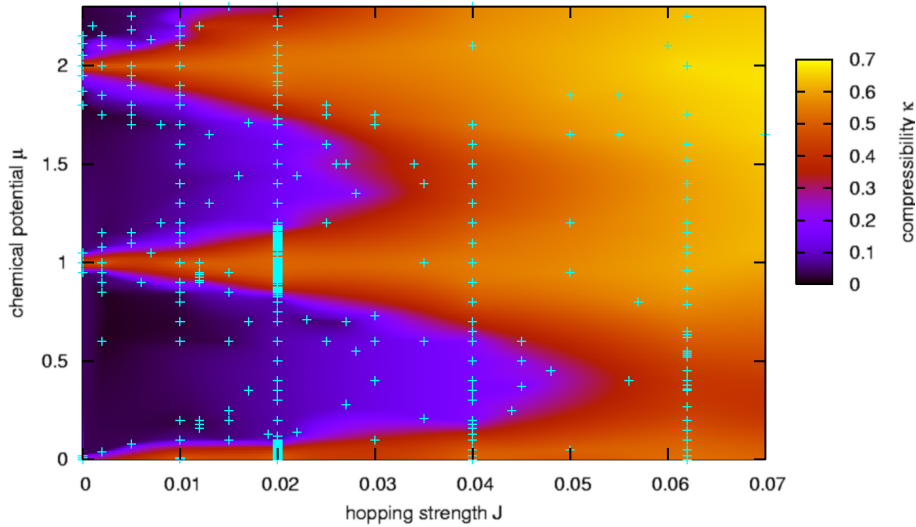}
\caption{Mean compressibility $\kappa$ of a $4\times 4$ lattice (with PBC) of Bose-Hubbard sites as function of the Hamiltonian parameters $J$ and $\mu$. One can clearly see the Mott insulator lobes ---characterised by low values (0 in the infinite case) of $\kappa$--- for densities $\rho=1,2,3$. and the surrounding superfluid phase. The crosses in cyan mark the parameter points for which a calculation was performed. To depict the values, an surface was interpolated between these points and is used to colour the plot. Unfortunatly, we could not fully get rid of artefacts due to the interpolation (see \ref{densityplotting}), and at those regions where the density of data points varies the contour lines become distorted. The resulting ``wobbling'' at regions with sparse data is hence not genuine and should vanish if one adds more data points. As cuts through the plane do not suffer from these presentation problems, we have calculated much more points for three fixed values of $J$ and show these cuts in Fig.\ \ref{BH_4x4_vertical_cuts}.}
\label{BH_4x4_phases}
\end{figure}

\begin{figure}
(a)\includegraphics[width=.42\textwidth]{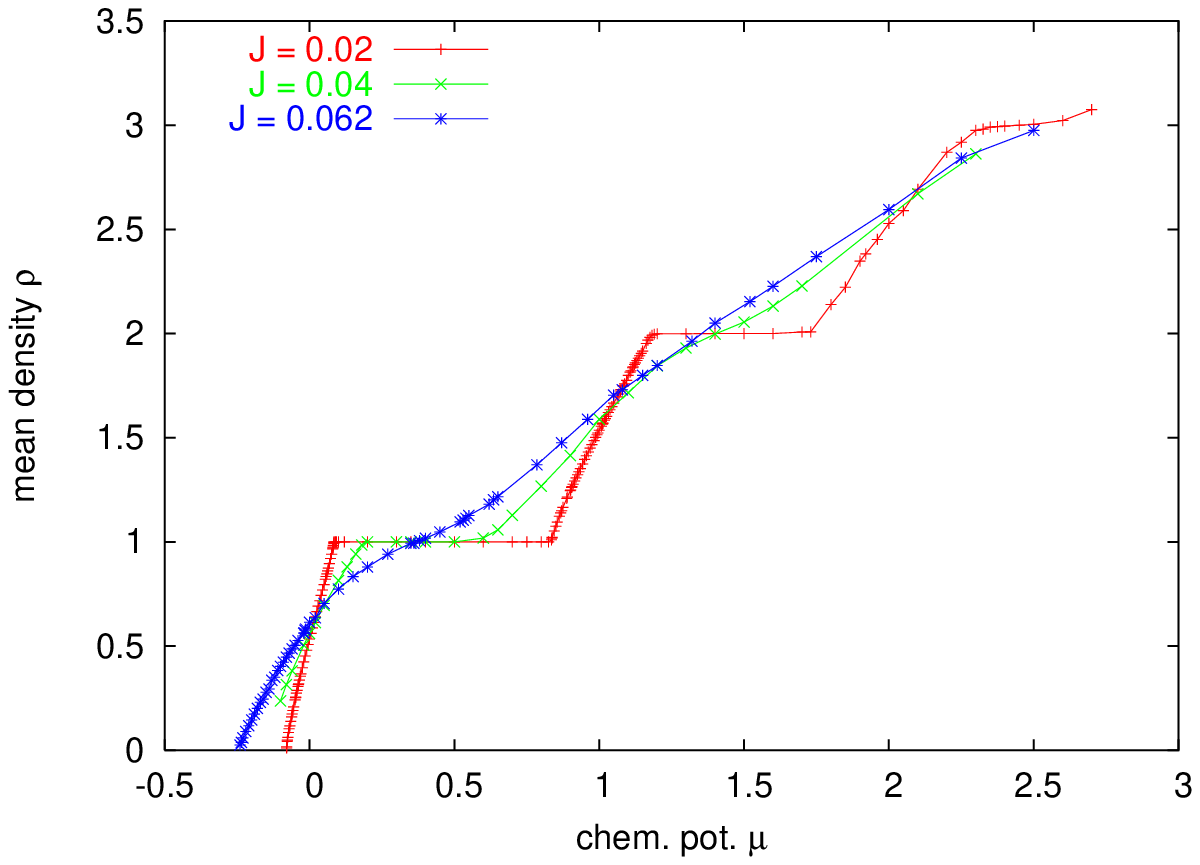} \quad
(b)\includegraphics[width=.42\textwidth]{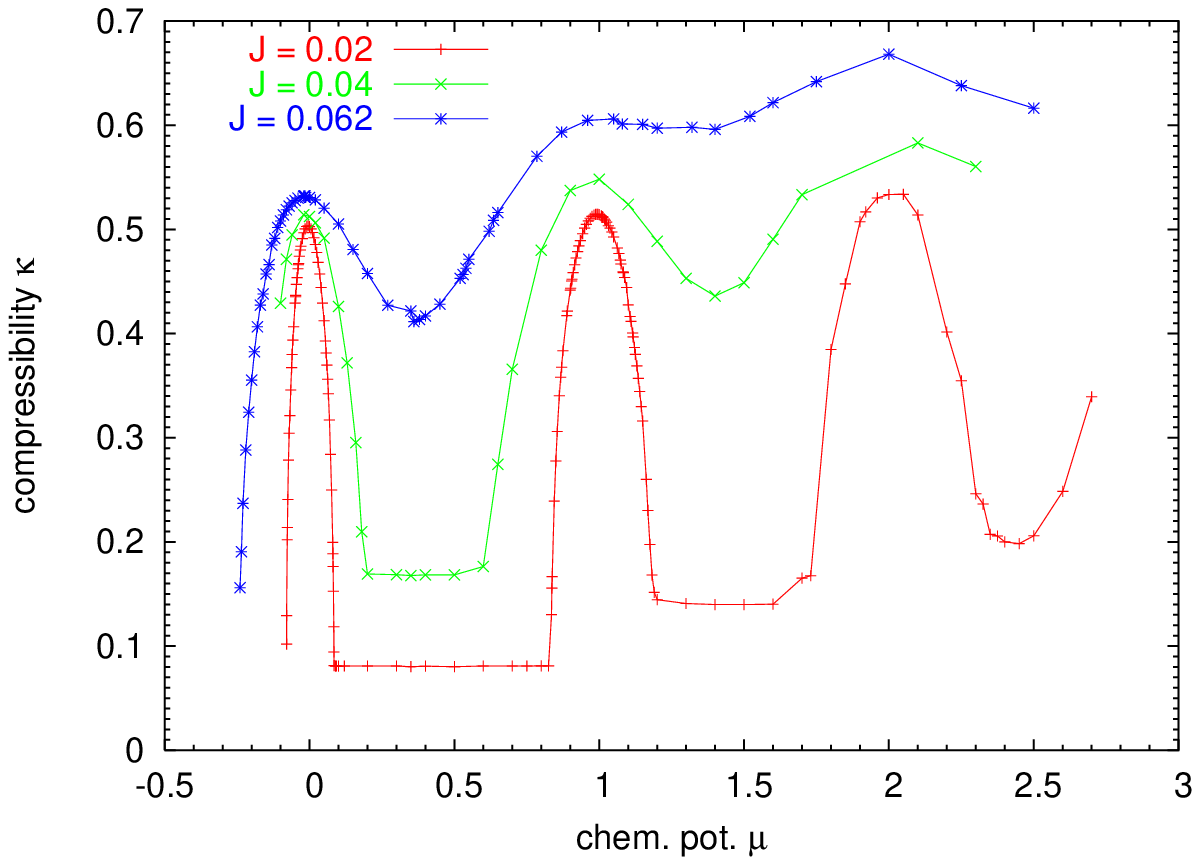}
\caption{As evident from the crosses in Fig.\ \ref{BH_4x4_phases}, we have many data point for $J=0.02, 0.04, 0.062$, which allow us to plot vertical cuts through the plane of Fig.\ \ref{BH_4x4_phases}. Here, we show the values of the average occupation number per site $\rho$ (a) and the compressibility $\kappa$ (b) for the mentioned three values of the hopping strengths $J$. The curve for $J=0.02$ looks smoothest because these points have been calculated to higher accuracy (cf.\ Fig.\ \ref{BH_4x4_density_density}).}
\label{BH_4x4_vertical_cuts}
\end{figure}

\begin{figure}
\includegraphics[width=.65\textwidth]{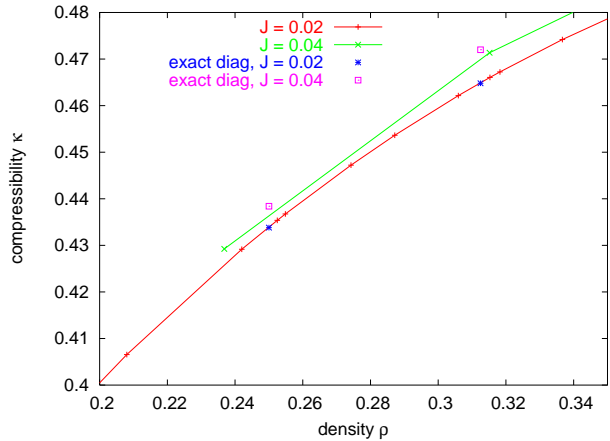}
\caption{For values of the density $\rho$ corresponding to very low (and integer) total particle numbers, we can obtain exact values for energy $E$ and compressibility $\kappa$ from diagonalisation of the full Hamiltonian. This shows that our method has good accuracy. (Note that the plot corresponds to a very small section of the steep left-most flank in Fig.\ \ref{BH_4x4_vertical_cuts}b.) For the $J=0.02$ curve, which has been calculated to especially high accuracy, the exact values for $\rho N=4,5$ (i.\,e., as $N=4^2$: $\rho=0.25, 0.325$) coincide with the approximated and interpolated values with an absolute deviation of only $10^{-4}$. Even for the values for $J=0.04$, which have been obtained with much fewer sweeping steps, the accuracy is below $2\cdot 10^{-3}$.}
\label{BH_4x4_comp_exact}
\end{figure}

\begin{figure}
\includegraphics[width=\textwidth]{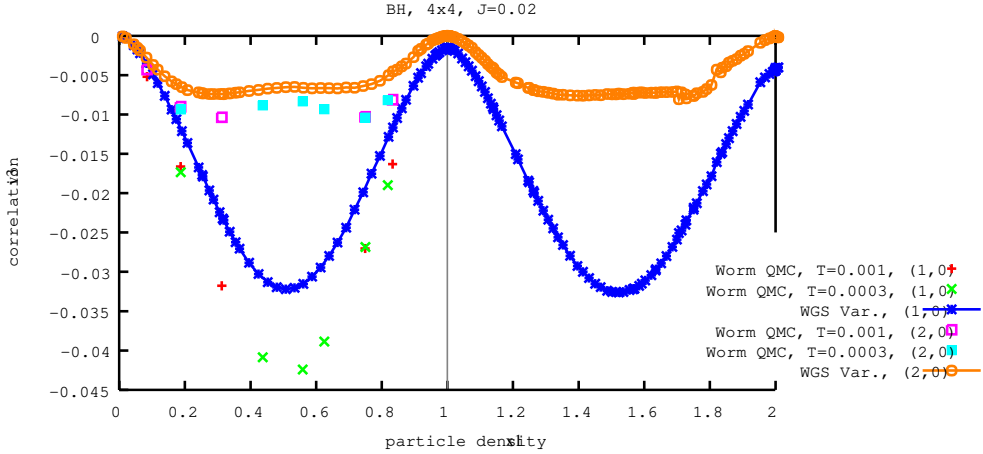}
\caption{Non-local properties of the ground state are much harder to obtain than local ones. In order to see whether our technique is also capable of yielding good results here, we calculated (for $J=0.02$ and varying values of $\rho$ for the 4x4 sites Bose-Hubbard system) the density-density correlation
$\gamma := \frac{1}{N}\sum_{a\in V} \left\langle {\hat n}_a {\hat n}_{a'} \right\rangle - \rho^2$ 
where $a'$ denotes the site that is one or two lattice step(s) to the right (denoted with (1,0) and (2,0). (Correlations of this kind have, incidentally, been studied recently in \cite{CakeStructure}.) The connected points show the results obtained with our approximation, the isolated points have been calculated using the worm code (quantum Monte Carlo technique)
\cite{Troyer_code} of the ALPS project \cite{ALPS} (at finite, but low temperature). The agreement is qualitatively ok but quantitatively not too precise. (Actually, the precision of two-point correlation is unfortunatly insufficient to obtain a good picture of the momentum distribution from their Fourier transformation.)}
\label{BH_4x4_density_density}
\end{figure}

\begin{figure}
\includegraphics[width=.7\textwidth]{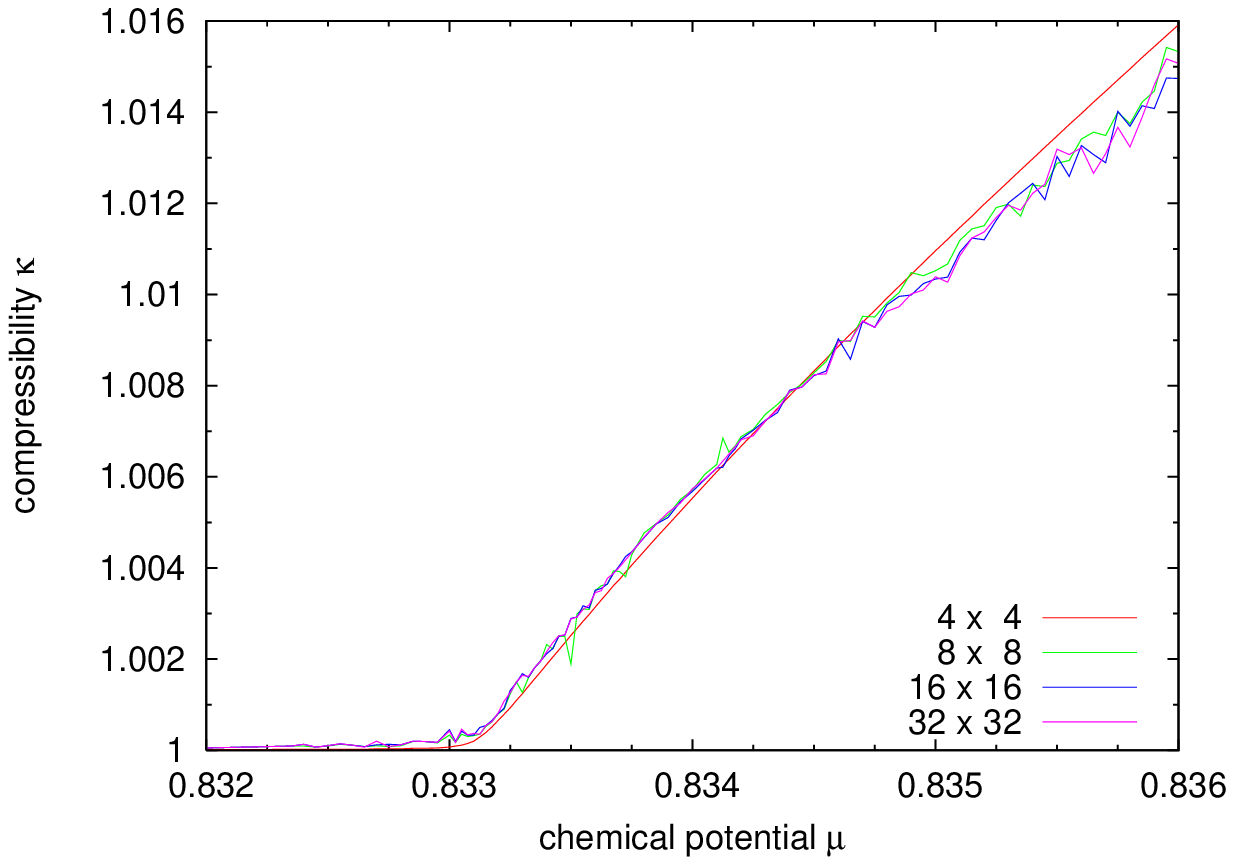}
\caption{Ground state approximation for Bose-Hubbard systems with up to $32\times 32$ sites. The figure shows the compressibility $\kappa$ as function of the chemical potential $\mu$ at the phase transition between the $n=1$ Mott lobe and the superfluid phase above of it (for a coupling of $J=0.02U$). As one can see, the numerics cope with the large amount of sites but fails to converge with a precision sufficient to make out clear finite-size scaling trends. The reason seems to be that the number of local minima increases very strongly with system size: While the use of sweeping technique allowed to get a smooth curve for the $4\times 4$ case, our attempts on the larger systems failed to get smoother than shown here. (We put considerably more effort in the data for $\mu<.8345$, but even there, the curves are still very noisy.)}
\label{larger}
\end{figure}

In order to simulate a bosonic system with our ansatz, we restrict the number of occupations at each site. For all the following calculations, we set the dimension of each site to $n=5$, i.\,e., the maximum occupation per site it $n-1=4$. The creation operator $\hat b^\dagger$ is defined such that $\hat b^\dagger\Ket{n-1}=0$ in order to truncate the Hilbert space. 

A good way to distinguish the Mott insulator from the superfluid phase it to look at the mean compressibility
\[\kappa=\frac{1}{N} \sum_{a\in V} \sqrt{\left( \left<{\hat n}_a^2\right> - \left<{\hat n}_a\right>^2 \right)},\]
which is strongly suppressed in the Mott insulator phase. Our results shows the form of the phase diagram in impressive clarity (Fig.\ \ref{BH_4x4_phases}). Although each data point only required a rather quick and rough calculation, one gets a good overview of the ground state properties in dependence of the Hamiltonian parameters $J$ and $\mu$. To better show the quantitative features, we have also plotted vertical cuts through the $(J,\mu)$ plane (Fig.\ \ref{BH_4x4_vertical_cuts}). Especially for the points at $J=0.02$, the zigzag sweeping technique (described later in Sec.\ \ref{sweeping}) was used to improve accuracy by more than an order of magnitude. This can also be seen from Fig.\ \ref{BH_4x4_comp_exact}: In this plot, we compare the observable $\kappa$, calculated for our approximand states, with exact values. To allow for this comparison, we have included values from exact Lanczos diagonalization. We are grateful to G.\ Pupillo, who supplied these numbers to us. He used a program, written for another project and using \textsc{Arpack} \cite{ARPACK}, that allows to diagonalize a small Bose-Hubbard system exactly if the number of particles is small as well. For a $4\times 4$ system, up to approx.\ 6 particles in the 16 sites can be treated. This corresponds to the very beginning of the plots of Fig.\ \ref{BH_4x4_vertical_cuts}, which we have magnified in Fig.\ \ref{BH_4x4_comp_exact}. The accuracy of $10^{-4}$ for the compressibility is competing well with the precision attainable with quantum Monte Carlo techniques.

While the compressibility is a local observable, the more challenging task is to study non-local observables such as density-density correlations of the form
\[ \gamma_{\nu} := \frac{1}{N}  \sum_{a\in V} \left[ \left< \hat n_a \hat n_{a'} \right> - \left(\sum_{a\in V} \left<\hat n_a\right>\right)^2\right],\]
where $a'$ is the site which has a fixed position relative to $a$, i.\,e. $\nu(a,a') =\nu(a)$. In Fig.\ \ref{BH_4x4_density_density}, we attempt this task for a $4\times 4$ lattice. Fig.\ \ref{larger} shows calculations for larger systems, up to $32\times 32$ sites. While the noise present in the latter plot is small on an absolut scale (note that the plot zooms in to a quite small parameter region) is is unfortunately still too large to prevent us from doing finite-size scaling.

\section{Performing the minimisation} \label{perf_min}

Usually, the Hamiltonian of a spin system is given in the form of a sum of terms each of which has support on only a small number of spins -- one or two in most physical cases. When the terms acting on single spins are absorbed into those acting on two spins, such a Hamiltonian can be written as 
\[ H = \sum_{(a,b)\in\mathcal{B}} H_{ab}^{(ab)}, \]
where $\mathcal{B}$ is the set of all pairs of spins, on which a term acts jointly. These pairs are called \textit{bonds} in the following, and they typically (but not necessarily) form a regular lattice. The \textit{bond Hamiltonians} $H_{ab}$ may all be equal or not, and only in the former case, the simplifications of Sec.\ \ref{phasesymms} can be used.

The minimisation problem \req{minprob} that we have to solve then takes the form
\[ E_{\rm min} = \on{min}_{\mathbf{x}\in\mathbb{R}^K} E(\mathbf{x});\quad \text{with } E(\mathbf{x}) = \sum_{(a,b)\in\mathcal{B}} \on{tr} H_{ab} \rho_{ab}\]

Finding a minimum of a general function of many parameters is a thoroughly researched but intrinsically hard problem. Our approach is described in the following. As we do not assume the reader's familiarity with numerical optimisation, we will explain some textbook knowledge.
\medskip

\subsection{Local search} \label{local_search}

Given a starting point $\mathbf{x}_0 \in \mathbb{R}^K$ in parameter space, the problem of \textit{local search} or \textit{local minimisation} is the task of finding a local minimum in the vicinity of $\mathbf{x}_0$. An exhaustive treatment of this topic can be found in the standard textbook \cite{NocedalWright} which covers all of the algorithms mentioned in the following in detail. In our case, we have to deal with \textit{unconstrained} (i.\,e., all values of the unbounded space $\mathbb{R}^K$ are admitted) \textit{nonlinear} (i.\,e., the energy function $E(\mathbf{x})$ does not have any simple structure that would allow the use of a more powerful, specialised algorithm) local minimisation. Algorithms for this case come in two classes: So-called \textit{direct methods} only require a means to evaluate the function at any given point, while \textit{gradient-based methods} also require a means to obtain the gradient $\bm{\nabla}_\mathbf{x}E(\mathbf{x})$ at any given point. 

Direct methods are convenient, but comparably slow. For very small systems (chains of up to 6 spin-1/2 sites, corresponding to less than 100 parameters), we could achieve convergence with direct methods, using the two most common ones, Nelder-Mead \cite{NelderMead} and Powell \cite{Powell_minimization} minimisation, with Powell minimization converging faster.

For any meaningful system size, however, direct methods are much too slow. Hence, we coded routines to obtain the derivatives of $E$ \wrt all kinds of parameters.\footnote{Note that it is not helpful to obtain the gradient by numerical differentiation, as this is hardly faster than using a direct method.} This required rather tedious calculation and coding, and the formulae and their derivation are summarised in \ref{gradient}.

Using the gradient functions, we tried the standard minimisation methods the literature offers, namely the Fletcher-Reeves conjugate-gradient method, the Polar-Ribi\`ere conjugate-gradient method and the Broyden-Fletcher-Goldfarb-Shanno (BFGS) method. We started using the implementations provided by the GNU Scientific Library \cite{GSL}, which, however, turned out to be not robust enough. Nevertheless, it could be established that convergence speed for our problem is as usually expected, i.\,e. Polar-Ribi\'ere (the oldest of the algorithms, from 1964) performs worst and BFGS (the newest, from the 1970s) performs best.

BFGS is a so-called quasi-Newton or Davidon algorithm. This means, it uses the gradients obtained at the points visited so far to build up an estimate to the Hessian of the function (assuming that the Hessian varies only slowly). The approximation to the Hessian (or more precisely, to the inverse of the Hessian) is then used to make a good guess for the next step. As the approximant is, like the Hessian, a $K\times K$ matrix, updating it at each step requires $\Or(K^2)$ steps, which scales worse than the calculation of $E(\mathbf{x})$, and hence, maintaining the BFGS data becomes more expensive than evaluating the function.

The textbook solution to this problem is to use the ``limited memory'' variant\footnote{The term ``limited memory'' shows that the problem of keeping the full matrix was then, in the 1980s, not so much seen in the time it takes to update the matrix but rather simple in the fact that a large matrix might not fit into the memory of the computer.} of BFGS, which is known as L-BFGS and stores only a list of the last, say 25, gradients, and uses this data to produce a Hessian approximant ``on the fly'' \cite{L-BFGS}. We have used the L-BFGS-B Fortran code \cite{L-BFGS-B}, which is very robust, not the least due to the excellent line search routine \cite{linesearchMoreThuente} that it uses.

A problem is the stop condition, which decides when convergence is assumed. We have tried several approaches: watching the norm of the gradient, the size of the steps, and the difference of the function value per step; these either taken point-wise, or averaged over the last, say 30, or 100, steps, or taken the maximum from the last 30-or-so steps. All this could not clearly predict convergence, as there seem to be long shallow slides, which tempt one to stop minimisation prematurely. In the end, we found that waiting until progress gets below machine precision is most viable. However, in the sweeping technique, described later, the minimzation can be stopped once it seems advantageous to first continue with a neighbor.
\medskip

The described technique only allows to find local minima. How do we find a good local minimum, or even the global one? Although the literature discusses many different heuristics and algorithms, it is a far from trivial task to find a good scheme. We have developed and tested two different heuristics, which shall be explained in the following two subsections. Both heuristics are two-phase methods \cite{review_twophase}, i.\,e., they combine a global driving scheme, that chooses points to start a local search from, with the local-search algorithm just discussed.

In both cases the minimization for a specific tuple of Hamiltonian parameters is typically performed several times. Whenever a new energy value is found which is lower than all values that have been found so far for this parameter tuple, the previous energy value and corresponding state is replaced by the new one. Hence, the longer one performs these heuristics the nearer one gets to the true ground state (or more precisely: to the lowest lying state within the variational class). We emphasize that we never discard a data point unless it has been ``underbid'' by a new calculation. This makes the result objective even if subjective judgement has been used in carrying out the heuristics.

\subsection{Multi-start and step-wise adding of superpositions}\label{multistart}

For the calculation of the results for the XY model (presented in Sec.\ \ref{XYmodel}), we tried  several different heuristics in order to ``move around'' local minima. We finally settled for the ``multi-start'' scheme described now, which turned out to work best, at least for the examples we studied: We started by choosing the parameters $\mathbf{x}_0$ for a state $\Ket{\Psi(\mathbf{x}_0)}$ with $m=1$ (i.\,e., without superpositions) uniformly at random from $[-5;5]^K$ and the used the L-BFGS algorithm (as explained in Sec.\ \ref{local_search}) to go downhill from there towards a minimum. We allowed this minimisation only to run for a limited number of function evaluations (typically a few hundred, or up to 1000), and then restarted with another randomly chosen initial point. Having done a number (say 15) of such ``trial runs'', the one that reached the lowest energy within the limited number of steps is kept, the other data is discarded. The best run is now allowed to continue for a significantly longer time, until the maximum number of ``main run'' steps (typically, several thousand function evaluation) is exhausted or the energy change falls below machine precision. Then, we increased the number $m$ of superpositions by one. This makes the parameter vector $\mathbf{x}$ longer, i.\,e. $2N+2$ real numbers have to be added (for $N$ complex deformation parameters and one complex superposition coefficient, cf.\ Eq.\ (\ref{num_param_unsymm}) for $n=2$). These are again chosen at random, but without changing the parameter values that have already been found. (It also helps to choose the new value for $\alpha_m$ with small modulus, such that the new parameters do not let the state stray to far from the already established good state.) Again, a number of trial runs is started, with different random numbers to extend the parameter vector, and the best one is allowed to continue for many more steps in the main run phase. This iterative extending of parameter values was looped until $m$ reached a certain value. This values does not have to be very large: for the results presented in Sec.\ \ref{XYmodel}, $m=3$ was sufficient.

A disadvantage of this heuristics is evident in Fig.\ \ref{1D_Ising}: Some points are much worse than their neighbours. For example, the point $B=1.09$ shows a sharp peak towards worse accuracy (orange line), while its neighbours to \textit{both} sides are better. What happened is that near the phase transition, the two phases compete to govern the ground state, and once the minimiser gets trapped by the catch-basin of one of the two phases it cannot switch to the other one. In most cases, the multi-start scheme will allow us to enter the main run within the catch-basin of the correct phase. If, however, the minimum energy of the two competing phases are very close, they cannot be distinguished during the short and rough trial runs, and it depends on mere chance to which phase we converge. The obvious solution is to use the value of neighbours which seem to have converged to lower energies as starting points in order to see if this allows to get to lower energies. This is the strategy that we tried next.

\subsection{Zigzag sweeping}\label{sweeping}

\begin{figure}
\includegraphics[width=\textwidth]{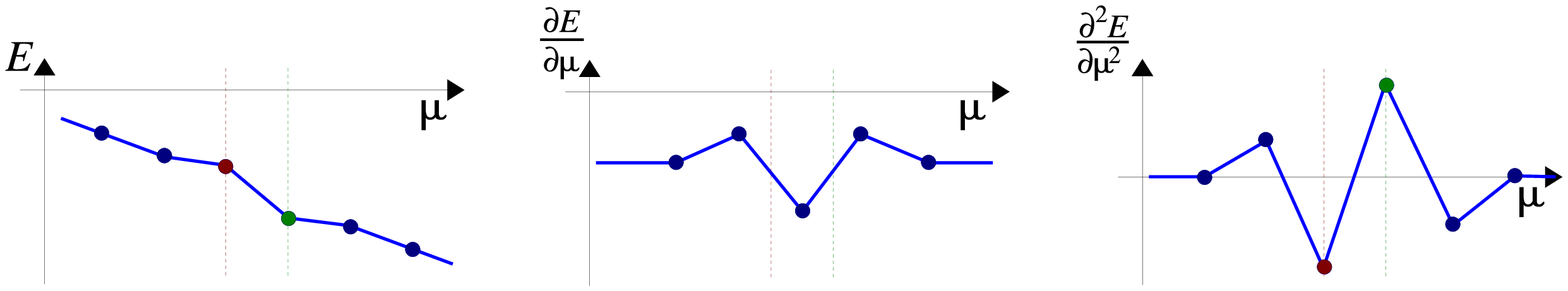}
\caption{Using derivatives to locally judge the quality of the approximant. Red (left) dashed vertical line: bad point, green (right) dashed vertical line: good point. For explanation see main text (Sec.\ \ref{sweeping}).}
\label{bad_point_good_point}
\end{figure}

All the results for the Bose-Hubbard model (as presented in Sec.\ \ref{BHmodel}) have been obtained without the use superpositions to the right of the phase gates, i.\,e., with $m=1$. Accuracy was instead improved in an iterative way using the following heuristics: Start minimisations from parameter vectors chosen at random for a variety of different values of Hamiltonian parameters (i.\,e., chemical potential $\mu$ and on-site repulsion $J$, for the Bose-Hubbard Hamiltonian) within the area of interest. Once the minimisations have converged more or less, compare each point with its neighbours. If one point looks better than a neighbouring, use this point's parameter vector to start a minimisation for the neighbouring point's Hamiltonian parameters.

In order to see how to do this in an objective way, look at the example of the red curve of Fig.\ \ref{BH_4x4_vertical_cuts}. There, $J=0.02$ is kept fixed and $\mu$ varies from -0.08 to 2.7. The data points are spaced rather closely ($\mu$ varies in steps of 0.003 up to 0.15). Hence, if we plot the energy $E$ versus the chemical potential $\mu$ and zoom in to look only at a few adjacent data points, we may expect to see simply a straight line. If the points lie close enough, any deviation from linearity is less likely for physical reasons but rather due to different quality (i.\,e., proximity to the global minimum) of the approximation at the points. Hence, we can interpret slight deviations towards higher [lower] energy as an indication that the point is a worse [better] approximant than its neighbours. As the slope varies too little to clearly see these differences it is helpful to take the second numerical derivative to enhance the differences. Following the sketch in Fig.\ \ref{bad_point_good_point}, a simple heuristic emerges: A pronounced peak in the second derivative means that the corresponding point is a better approximant than its neighbours. Hence, use its parameter vector as initial values to redo the minimisation at the neighbouring Hamiltonian parameters. If one of the two neighbours has a much lower second derivative than the other, redo only this one. Conversely, a point with a pronounced dip in the second derivatives should be re-done, starting with the parameter vector of one of its neighbours, normally the one with the higher second derivative.

Close to a phase transition, the procedure may get stuck because the step from one neighbour to the next changes the state too much. In this case, one should insert a new data points between the point that failed to get better and the neighbouring point used for the initial value.

\subsection{Outlook to other minimization techniques} \label{min_outlook}

The literature on unconstrained nonlinear minimization is vast, and finding a good global minimization scheme requires a lot of trial and error. Apart from the two-phase heuristics described above, we have also tried genuine global minimization techniques, namely simulated annealing \cite{simulated_annealing} and differential evolution \cite{diffevol}. Both are genuinly global in the sense that they do not employ a local search stage. However, they thus cannot take advantage of the possibility to calculate the gradient. Hence, it is not surprising that simulated annealing converged much too slowly to be of use. (Simulated annealing is used in many different fields with much success but usually for functions with a convoluted potential surface but only few variables. We have several hundreds or even thousands of variables.) Differential evolution is a genetic algorithm and shows the ---on first sight surprising--- feature of converging to the mean field solution. (This seems explicable from the fact that crossing two genotypes in different basins has to end up in a ``compromise'', which is mean field.)

One further possibility might be basin hopping, which is a family of techniques (reviewed in \cite{review_basin_hopping}) that combine simulated annealing with a local search phase in order to overcome the problem states in the previous paragraph. These ideas are quite recent and research is still ongoing. So far, however, it seems that the basin hopping requires to perform very many local searches which hence have to converge fast. This is unfortunately not so in our case. It seems conceivable that variants can be developped that only use rough and hence fast local searches, and this might be a way to proceed with our method.

Another ansatz is using a clustering stage in the global phase of a two-phase method \cite{clustering_globopt}. This allows to make multi-start much more efficient but has two difficult requirements: (i) One needs to factor any degenerecies in the minima out of the parameter space. We have not yet studied whether this is possible. (ii) The number of local minima must be small enough that one has a decent chance to encounter all of them during the local searches. Unfortunately, especially the calculations for Fig.\ \ref{larger} have brought us to the observation that the number of minima seems to grow very fast with the system size.

A further technique that we have tried is imaginary time evolution, which works as follows. Given an initial state $\Ket{\Psi(\mathbf{x}_0)}$ chosen at random, we can find an estimate $\Ket{\Psi(\mathbf{x}_{i+1})} \approx \on{nrm} e^{-\Delta t H} \Ket{\Psi(\mathbf{x}_{i+1})}$ for the discretized evolution of the state under the system Hamiltonian in the imaginary time direction. As for most initial states $\Ket{\Psi_0}$, $\Psi_{\rm G} = \lim_{t\to\infty} \on{nrm} e^{-t H}$ is the ground state, this iterative evolution should converge to a good approximation of the ground state. We have decomposed $e^{-\Delta t H}$ into a product of bond terms $e^{-\Delta t H_{ab}}$ using standard Trotter decomposition and then tried to find the $\Delta\mathbf{x}\in\mathbb{R}^K$ that maximizes the overlap
\[\left|\frac{\Braket{\Psi(\mathbf{x}+\Delta\mathbf{x})|e^{-\Delta t H_{ab}}|
\Psi(\mathbf{x})}}{\sqrt{\Braket{\Psi(\mathbf{x}+\Delta\mathbf{x})|\Psi(\mathbf{x}+\Delta\mathbf{x})} \Braket{\Psi(\mathbf{x})|e^{-2\Delta t H_{ab}}|\Psi(\mathbf{x})}}}\right|.\]
Unfortunatly, the maximization failed to give good result even for arbitrarily small time steps $\Delta t$ and we thus abandoned this approach.

We should also mention that for the some of the results of or previous paper \cite{WGSMinPRL}, we (actually, M.\ Plenio, who programmed this part) have used a Rayleigh minimization technique: One restricts the energy function $E(\mathbf{x})$ in the sense that one keeps all but a few parameters fixed. For certain such subsets of only a few parameters, namely for the set of parameters corresponding to a single local unitary or to the phases and deformations for one pair of qubits, one can write the restricted energy function as quotient of two quadratic forms. This is also known as a generalized Rayleigh quotient and the global minimum can be find via a generalized eigenvalue problem. Such a ``global minimum'' typically is, however, not even a local minimum of the full energy function. The reason that we got good result for the Ising model in \cite{WGSMinPRL} seems now, in retrospect, have been due to the extraordinarily benign form of the corresponding energy landscape. Hence and because the scheme cannot easily generalized to spins higher than 1/2, we did not persue this any further.

\section{Conclusion and outlook} \label{conclusion_and_outlook}

To conclude, we have presented a class of variational states that holds promise to approximate the ground states of spin systems and bosonic systems. The advantageous properties of this class is that it includes states with an arbitrarily high entanglement and the possibility to adapt to arbitrary geometries and number of spatial dimensions. We have shown how to calculate expectation values of observables for these states and demonstrated the approximation of the ground state for two model systems, namely the XY spin-1/2 model and the Bose-Hubbard model, in one and two dimensions. Furthermore, we have explained heuristics suitable to drive the minimization.

The method works for small systems and maps out the rough structure of phase diagrams. (The system sizes, though small, were sufficient to see the phase boundaries even though phase boundaries are defined, strictly speaking, only in the thermodynamical limit.) We can calculate observables for states in systems of considerable size but have problems in approximating the ground state in larger system to precision sufficient to see actual differences between different system sizes and hence to do finite-size scaling studies.

It seems likely that this is not because there were no states in our variational class which were close enough to approximate such ground states well. Rather, we simply cannot find them because our minimzation gets trapped in local minima. Can the avoid this? This is the crucial question for the future development of the scheme, and at this moment, we may only offer some thoughts on that: It seems unlikely that the choice of another generic global minimzation algorithm is able to steer around these local minima better that those algorithms that we have tried. For further progress, it seems hence desirable to have a better understanding of the shape and structure of the manifold $\Psi(\mathbb{R}^K)\subset\mathcal{H}$, i.\,e., our variational set of states as described by the mapping from the parameter space. Is, for example, this manifold ``folded'' more and curved stronger than the equi-energy surfaces of typical system Hamiltonians? This might explain, why there are so many local minima --- and getting a better grasp on topology and metric of the mapping $\Psi$ and its image could be most helpful in finding a better way to steer towards good minima.

\ack

We are indebted to G.\ Pupillo for providing the data from exact diagonalisation used in Fig.\ \ref{BH_4x4_comp_exact}, and we would like to thank G.\ Pupillo,   H.-P.\ B\"uchler and J.\ Eisert for helpful explanations on the Bose-Hubbard model. We thank M.~B.\ Plenio and V.\ Verstraete for discussions. The numerical calculations presented in this work have been carried out using the compute cluster of the University of Innsbruck's Konsortium Hochleistungsrechnen. This work was supported by the Austrian Science Foundation (FWF) and the European Union (Projects OLAQUI and SCALA). 

\appendix

\section{Notes on the implementation} \label{implementation}

\subsection{Avoiding overflows}

A certain detail is worth mentioning as it may cause some difficulty in the implementation: As the product (\ref{rho_compact}) contains $\Or(N)$ terms, its values grows exponential with the system size $N$. Even for factors which are quite close to 1, the value will leave the range of floating-point arithmetics (on most computers, ca.\ $10^{-308}\dots 10^{308}$) for even moderate values of $N$. To avoid this, one has to compute the product by summing up the \textit{logarithms} of the matrix elements of $\rho_{ab,c}^{jk}$, then subtracting a constant from this sum, and then exponentiating the result component-wise. The subtraction of the constant does not change the final result, as it formally cancels against the final normalisation to unit trace. The exact value of the constant is hence irrelevant, but it has to be chosen large enough to avoid a floating-point overflow during exponentiation, but not so large that the elements of all the matrices $\rho_{ab}^{jk}$ vanish due to floating-point underflows. (That some elements of some of the matrices $\rho_{ab}^{jk}$ suffer an underflow is, however, unavoidable, but harmless, as their contribution to the result is evidently insignificant.) Especially for large systems, the constant has to be read\-justed during the minimization.

\subsection{Choice of programming languages}

We have written two implementations of our algorithm. The first one, called ``ewgs'' is specialised for spin-1/2. It was used for the results on the XY model (Sec.\ \ref{XYmodel}), and also for the results presented in \cite{WGSMinPRL}.\footnote{For completeness, we should point out one difference between the description in this article and the implementation: In ``ewgs'', the unitaries are not parametrised using the Cayley transform, but rather as linear combination of the identity and the Pauli matrices: $U=\frac{u_0\mathbb{1} + u_1\sigma_x + u_2\sigma_y + u_3\sigma_z}{u_0^2+u_1^2+u_2^2+u_3^2}$.} The other, more recent program is called ``hwgs'' and may be used for spins of any size $n$. ``ewgs'' is mainly written in C++, only the outer drivers are written in Python. Python \cite{PythonHL} is a very modern, quite powerful scripting language, that features high-performance just-in-time compilation, an exceptionally comprehensive low- and high-level library, an open-source license and excellent inter-platform portability. The development of a numerics library for Python has reached maturity quite recently with the release of NumPy \cite{NumPyURL,NumPyGuide}. Due to the higher level of the language, development is much faster in Python than in C++. This makes it advisable to do most of the coding in Python and only write the ``hot spots'', i.\,e., the proverbial 10\% of the code in which the processor spends 90\% of the time, in an optimizing compiled language such as C++. This approach, though it may sound unusual to a traditionally oriented computer physicist, has been used in several places with much success (see e.\,g. the advocacy in \cite{Python_for_linalg}), and from our experiences, we clearly recommend its use. Hence, for our second implementation, ``hwgs'', we followed this paradigm consequently and wrote only a small part in C++. This part was bound to the main Python code using SWIG \cite{SWIG}. For the local minimizer we used in both implementations the L-BFGS-B Fortran code \cite{L-BFGS-B}, linked to Python with the help of the tool f2py \cite{f2py}.

\subsection{Performance}
\begin{figure}
\includegraphics[width=.6\textwidth]{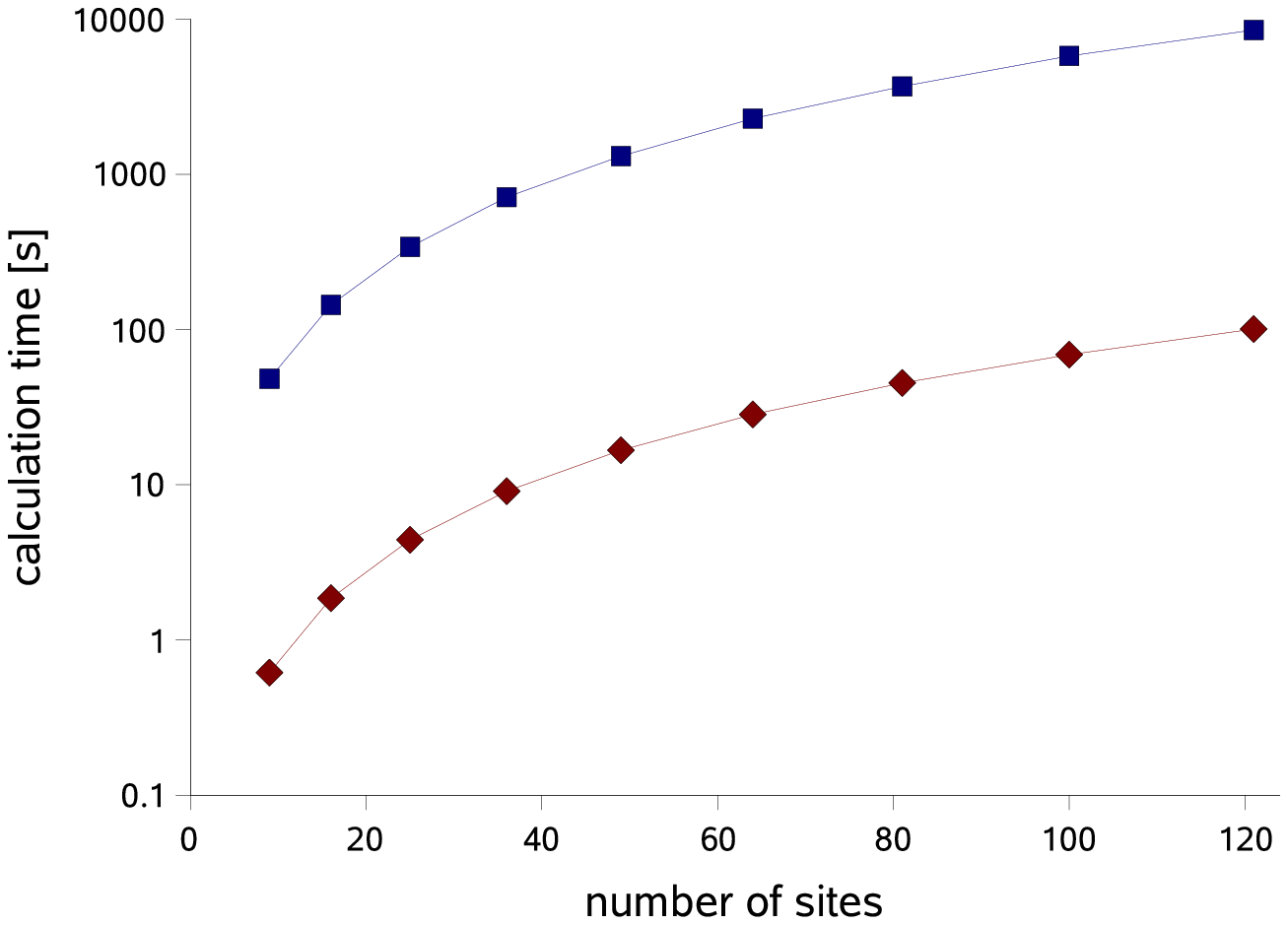}
\caption{Performance of our ``hwgs'' implementation: For a Bose-Hubbard system on a square lattice of varying size, the calculation time for a single reduced density matrix of a pair of sites (red diamonds) and for the full gradient of the energy (derivatives \wrt \textit{all} parameters) (blue squares) is shown. These calculations have been done for states with $n=5$ levels per site and no superpositions ($m=1$). The program was run on AMD Opteron machines clocked at 2.2 GHz.}
\label{hwgs_perf}
\end{figure}

The performance of the ``hwgs'' implementation can be seen in Fig.\ \ref{hwgs_perf}. The blue curves shows the time required to calculate energy and full gradient for one parameter vector at various system sizes. In order to see the time required to find a good approximand, this has to be multiplied with the number of function evaluations needed by the minimiser. 

Usually, one wants to find approximands for several different values of the Hamiltonian parameters. Then, one can save much time by running these minimisation in parallel if one has access to a computer cluster.

\subsection{Availability}
We would welcome to see our code been used in further projects. Hence, researchers who are interested in applying our code in their own projects are encouraged to contact the authors.

\subsection{Density plots} \label{densityplotting}

The plotting technique used to obtain Figs.\ \ref{XY1DParamPlot} and \ref{BH_4x4_phases} merits a brief explanation. For these plots, we calculated the plotted quantity at different value pairs for the quantities at the $x$ and $y$ axes. In order to work out interesting feature, we did not evaluate at a fixed grid but rather started with some losely spaced points to get an overview and then added more and more points at regions with interesting features. This allowed us to ``explore'' the parameter plane. However, it leaves us with a list of data points at irregular positions, which makes the usual 3D mesh plots unsuitable (as a mesh plot requires data from a regular grid). This is why we visualize the data instead with density plots, using colour to indicate $z$ height. To obtain the colours we interpolated between the data point, and for this, we experimented with two interpolation algorithms, namely Akima's spline method \cite{Akima3D} and the Sibson's natural neighbours method \cite{Sibson_interpolation}. As the former has problems with strongly varying curvature (and this is the case here: the data varies more strongly near the phase transition than in the interiours of the phases) we used Sibson's method and produced Figs.\ \ref{XY1DParamPlot} and \ref{BH_4x4_phases} with the help of the \textsc{Natgrid} implementation \cite{NatGrid} of Sibson's algorithm.

\section{Calculating the gradient of the energy with respect to the parameter vector} \label{gradient}

For use in the gradient-based minimisation we need a fast way to obtain the gradient of the energy function $E(\mathbf{x})$. For the following, we assume that the Hamiltonian can be written in bond form,
\[ H = \sum_{(a,b)\in\mathcal{B}} H_{ab}^{(ab)}.\]
As before, $\mathcal{B}$ is the set of bonds, i.\,e. of pairs of interacting spins. In many cases $H_{ab}$ is the same for all bonds $ab$, but having an inhomogeneous Hamiltonian is no complication.

As the energy function is given by $E(\mathbf{x}) = \sum_{(a,b)\in\mathcal{B}} \tr H_{ab}\rho_{(a,b)}$, its gradient consists of a sum of derivatives of the reduced density matrices
\[ \frac{\partial E(\mathbf{x})}{\partial x_l} = \sum_{(a,b)\in\mathcal{B}}\tr H_{ab} \frac{\partial \rho_{ab}}{\partial x_l}. \]

We shall now derive formulae for the components of the gradient, i.\,e., for the derivatives \wrt the different kinds of parameters.

\subsection{Derivatives \wrt the parameters for the local unitaries}

The derivative of a matrix exponential \wrt the components of the exponentiated matrix (or of linear combinations of these) is a very involved problem. Not only is the integral representation of this parametric derivative, though simple, in no way obvious, but also is the evaluation of this integral a very non-trivial matter. For a review of the history of this problem and current state of knowledge, consult Ref.\ \cite{deriv_matr_exp}.

For us, this is the main reason why we do not use the exponentiation of a Hermitian matrix for the parametrisation of the local unitaries, but rather the Cayley transform of it, for the latter involves only a matrix inverse, whose parametric derivative is expressed by a simple formula: For any invertible square matrix $A=A(t)$ that depends differentiably on a real parameter $t$, we have
\bEq \frac{\mathrm{d}A^{-1}}{\mathrm{d}t} = -A^{-1}\frac{\mathrm{d}A}{\mathrm{d}t} A^{-1} \label{deriv_inv_mat} \eEq
(for a proof, see e.\,g.\ \cite{deriv_inv_matr}). 

We need $n^2$ real parameters to parametrise a Hermitian $n\times n$ matrix, which we arrange to form a $n\times n$ upper triangular matrix $\tilde A$ with real entries in the diagonal, complex entries in the upper triangle and zeroes in the lower triangle. $A=\tilde A + \tilde A^\dagger$ is now Hermitian and \bEq U=\left(i\mathbb{1} + \tilde A + \tilde A^\dagger\right) \left(i\mathbb{1} - \tilde A - \tilde A^\dagger\right)^{-1} \label{Cayley_trafoB}\eEq
is unitary.
Using \req{deriv_inv_mat}, we get
\begin{eqnarray} 
\fl\frac{\partial U}{\partial \ReIm A_{kl}} &= \OneI
(\mathbb{1}+U) \left(\Ket{k}\Bra{l} \PlusMinus \Ket{l}\Bra{k}\right)(i\mathbb{1}-A)^{-1},\label{derivCayley}\\
\fl\frac{\partial U^\dagger}{\partial \ReIm A_{kl}} &= - \OneI
(i\mathbb{1}+A)^{-1} \left(\Ket{k}\Bra{l} \PlusMinus \Ket{l}\Bra{k}\right)(\mathbb{1}+U^\dagger)
=\left(\frac{\partial U}{\partial \ReIm A_{kl}}\right)^\dagger.\nonumber
\end{eqnarray}

We use this to calculate
\begin{eqnarray} \fl\frac{\partial E}{\partial A_{a,kl}} &= \sum_{b:(b,a)\in\mathcal{B}} \tr H_{ba} \Bigg[ 
\left(U_b\otimes\frac{\partial U_a}{\partial A_{a,kl}}\right) 
\tilde{\tilde{\rho}}_{ba}
\left(U_b\otimes U_a\right)^\dagger + \hfill \nonumber \\ 
\fl & \hspace{10em} + 
\left(U_b\otimes U_a\right) \tilde{\tilde{\rho}}_{ba}  
\left(U_b\otimes\frac{\partial U_a}{\partial A_{a,kl}}\right)^\dagger
\Bigg] + \nonumber \\
\fl &+ \sum_{c:(a,c)\in\mathcal{B}} \tr H_{ac} \Bigg[ 
\left(\frac{\partial U_a}{\partial A_{a,kl}}\otimes U_c\right) \tilde{\tilde{\rho}}_{ac} 
\left(U_a\otimes U_c\right)^\dagger + \nonumber \\
\fl &  \hspace{10em} +
\left(U_a\otimes U_c\right) \tilde{\tilde{\rho}}_{ac}  
\left(\frac{\partial U_a}{\partial A_{a,kl}}\otimes U_c\right)^\dagger
\Bigg], \nonumber 
\end{eqnarray}
where $\tilde{\tilde{\rho}}_{bc}$ is the reduced density matrix without application of the local unitaries, i.\,e.\ $\tilde{\tilde{\rho}}_{bc} = (U_b\otimes U_c)^\dagger \rho_{bc} (U_b\otimes U_c)$.

\subsection{Derivatives \wrt the deformation parameters} \label{deriv_deform_S}

For the derivatives \wrt the parameters $\on{Re} d_{c,s}^l$ and $\on{Im} d_{c,s}^l$, we have to take care of the normalisation of $\rho_{ab}$ as it depends on those parameters. We abbreviate the middle line of \req{rho_ab_S} with $\tilde\rho_{ab}$ and start with using \req{rho_with_Dcheck} in order to see that
\[\tilde\rho_{ab} = \sum_{j,k=1}^m \alpha_j \alpha_k^* \check D_{ab}^{jk} \odot \rho_{ab}^{jk}\]
where (according to \req{def_Ket_D_A})
\[ \check D_{ab}^{jk} := \sum_{\mathbf{r},\mathbf{r}'\in \mathbb{S}} d_{a,r_1}^a d_{a,r_2}^b {d_{a,r'_1}^a\!\!\!}^*\, {d_{a,r'_2}^b\!\!\!}^*\, \Ket{\mathbf{r}}\Bra{\mathbf{r}'}. \]

We write the derivative as
\bEq \fl\frac{\partial \rho_{ab}}{\partial \ReIm d_{c,s}^l} = (U_a\otimes U_b)W_{\phi_{ab}}
\left( \frac{\partial}{\partial \ReIm d_{c,s}^l}
\frac{\tilde\rho_{ab}}{\tr\tilde\rho_{ab}}
\right) 
W_{-\phi_{ab}} (U_a\otimes U_b)^\dagger, \label{deriv_deform_outer1}\eEq
where the term in parentheses becomes
\bEq \frac{\frac{\partial\tilde\rho_{ab}}{\partial\ReIm d_{c,s}^l} \tr\tilde\rho_{ab} - 
\tilde\rho_{ab} \tr \frac{\partial\tilde\rho_{ab}}{\partial\ReIm d_{c,s}^l}}
{\left(\tr\tilde\rho_{ab}\right)^2} \label{deriv_deform_outer2} \eEq

In order to evaluate $\partial\tilde\rho_{ab}/\partial\ReIm d_c^l$, we distinguish three cases, namely (i) $c=a$, (ii) $c=b$, (iii) $c\notin\{a,b\}$.

\textit{Case (i):} The only term in the middle line of Eq.\ (\ref{rho_ab_S}) that depends on $d_a^l$ is $\check D_{ab}^{jk}$ and this only for those terms in the sum, where $j=l$ or $k=l$. One finds
\[ \fl \frac{\partial\tilde\rho_{ab}}{\partial\ReIm d_{as}^l} = 
\OneI \alpha_l \sum_{k=1}^m \alpha_k^* 
\sum_{r_1,r_1',r_2'\in\mathbb{S}} d_{br_1}^l {d_{ar_1'}^k\!\!\!}^*\, {d_{br_2'}^k\!\!\!}^*\,
\Ket{r_1s}\Bra{r_1'r_2'} \odot \rho_{ab}^{lk} + \text{H.c.} \]

\textit{Case (ii):} Analogous:
\[ \fl \frac{\partial\tilde\rho_{ab}}{\partial\ReIm d_{bs}^l} = 
\OneI \alpha_l \sum_{k=1}^m \alpha_k^* 
\sum_{r_2,r_1',r_2'\in\mathbb{S}} d_{ar_2}^l {d_{ar_1'}^k\!\!\!}^*\, {d_{br_2'}^k\!\!\!}^*\,
\Ket{sr_2}\Bra{r_1'r_2'} \odot \rho_{ab}^{lk} + \text{H.c.}, \]

\textit{Case (iii):} For $c\notin\{a,b\}$, $\check D_{ab}^{jk}$ is independent of $d_c^l$, but $\rho_{ab}^{jk}$ is now dependent. We get:
\begin{eqnarray} 
\fl \frac{\partial\tilde\rho_{ab}}{\partial\ReIm d_{cs}^l} = 
\OneI \alpha_l \sum_{k=1}^m \alpha_k^* 
\check D_{ab}^{lk} \odot \\
\fl\qquad{}\odot {d_{cs}^k\!\!}^*\,
\left( \exp \rmi\left(
\Phi_{ac}^{r_1s}-\Phi_{ac}^{r_1's}+\Phi_{bc}^{r_2s}-\Phi_{bc}^{r_2's}
\right)\right)_{\mathbf{r},\mathbf{r}'\in\mathbb{S}^2} \odot
\bigodot_{e\in V\backslash\{a,b,c\}} \rho_e^{lk} +\text{H.c.},
\nonumber\end{eqnarray}
where $\rho_e^{lk}$ is given by \req{rho_c_jk_S}.

\subsection{Derivatives \wrt the superposition coefficients}

The derivatives \wrt $\on{Re}\alpha_l$ and $\on{Im}\alpha_l$ are found the same way as for the deformation, and one gets
\[
\frac{\partial\on{nrm}\tilde\rho_{ab}}{\partial\alpha_{ab}} = 
\frac{\frac{\partial\tilde\rho_{ab}}{\partial\ReIm \alpha^l} \tr\tilde\rho_{ab} - 
\tilde\rho_{ab} \tr \frac{\partial\tilde\rho_{ab}}{\partial\ReIm \alpha_l}}
{\left(\tr\tilde\rho_{ab}\right)^2}
\]
with
\[\frac{\partial\tilde\rho_{ab}}{\partial \ReIm \alpha_l} =
\OneI \sum_{k=1}^m \alpha_k^*
\Ket{D^l_{(a,b)}}\Bra{D^k_{(a,b)}} \odot 
\tilde\rho_{ab}^{lk} + \text{H.c.},
\]
where $\Ket{D^l_{(a,b)}}$ is defined in \req{def_Ket_D_A}.

\subsection{Derivatives \wrt the phases}

We first introduce
\[ \check W_\Phi = \Ket{\Phi}\Bra{\Phi} \quad\text{with } \Ket{\Phi} = \sum_{\mathbf{r}\in\mathbb{S}^2}\Ket{\mathbf{r}} \rme^{\rmi\Phi^{r_1r_2}}, \]
so that we can write ---due to \req{diag_to_hadm_prod}--- \req{rho_ab_S} as
\[ \rho_{ab} = (U_a\otimes U_b) \left(\check W_{\Phi_{ab}}\odot \tilde\rho_{ab}\right)  (U_a \otimes U_b). \]

We have to take care that in case of a symmetrisation according to Sec.\ \ref{phasesymms} a phase can occur more than once in the expression for $\rho_{ab}$, and hence, we make use of the phase index mapping $\nu(a,b)\in\{1,\dots,R\}$ (Sec.\ \ref{phasesymms}), that associates with every pair of spins $a$, $b$ a phase matrix $\Phi_{\nu(a,b)}$. We write (with $\mathbf{r}\equiv(r_1r_2)$)
\[ \fl \frac{\partial \rho_{ab}}{\partial \Phi_{\nu}^{\mathbf{r}}} =
\frac{1}{\tr\rho_{ab}} (U_a\otimes U_b) \left(\frac{\partial\check W_{\Phi_{\nu(a,b)}}}{\partial\Phi_{\nu}^{\mathbf{r}}} \odot \tilde\rho_{ab} +
\check W_{\Phi_{\nu(a,b)}} \odot \frac{\tilde\rho_{ab}}{\partial\Phi_{\nu}^{\mathbf{r}}}\right)(U_a\otimes U_b)^\dagger
\]
and proceed to discuss the two derivatives in this expression.

The first one is evidently non-zero only if $\nu=\nu(a,b)$ and then evaluates to
\[ \frac{\partial\check W_{\Phi_{\nu(a,b)}}}{\partial \Phi_{\nu(a,b)}^{\mathbf{r}}} = i \check W_{\Phi_{\nu(a,b)}} \odot \left(
\delta_{\{q_1,q_2\},\{r_1,r_2\}} - \delta_{\{q_1',q_2'\},\{r_1,r_2\}} \right)_{\mathbf{q},\mathbf{q}'\in\mathbb{S}^2}. \]
The set notation in the Kronecker deltas accounts for the fact that $\Phi$ is symmetrised, $\Phi^{r_1r_2} = \Phi^{r_2r_1}$, (cf. again Sec.\ \ref{phasesymms}) and hence, the order of the components of the vectors $\mathbf{r}$, $\mathbf{q}$, and $\mathbf{q}'$ must be disregarded.

For the second term, we pull the derivative inwards
\[\frac{\partial\tilde\rho_{ab}}{\partial\Phi_\nu^\mathbf{r}} =
\sum_{j,k=1}^m \alpha_j \alpha_k^* \check D_{ab}^{jk} \odot \frac{\partial\rho_{ab}^{jk}}{\partial\Phi_\nu^\mathbf{r}},\]
then rewrite \req{rho_compact} as
\[ \rho_{ab}^{jk} = \bigodot_{c\in V\backslash\{a,b\}} \sum_{s\in\mathbb{S}} d_{cs}^j {d_{cs}^k}^* \Ket{\tilde\Phi_{abc}^s}\Bra{\tilde\Phi_{abc}^s} \] 
with
\[ \Ket{\tilde\Phi_{abc}^s} = \sum_{q_1,q_2\in\mathbb{S}} \Ket{q_1q_2} 
\exp \left[\rmi\left(\Phi_{\nu(a,c)}^{q_1s} + \Phi_{\nu(b,c)}^{q_2s}\right)\right] \]
and continue
\begin{eqnarray}\frac{\partial\rho_{ab}^{jk}}{\partial\Phi_\nu^\mathbf{r}} = 
\sum_{e\in V\backslash\{a,b\}} \left(
\bigodot_{c\in V\backslash\{a,b,e\}} \sum_{s\in\mathbb{S}} d_{cs}^j {d_{cs}^k}^* \Ket{\tilde\Phi_{abc}^s}\Bra{\tilde\Phi_{abc}^s} \right) \odot\nonumber\\
\qquad\qquad{}\odot
\sum_{s\in\mathbb{S}} d_{es}^j {d_{es}^k}^* \frac{\partial \Ket{\tilde\Phi_{abe}^s}\Bra{\tilde\Phi_{abe}^s}}{\partial\Phi_\nu^\mathbf{r}}.\nonumber
\end{eqnarray}
The sum over $e$ formally runs over $N-2$ terms. Most of these vanish, however, namely all those for which neither $\nu=\nu(a,e)$ nor $\nu=\nu(b,e)$. For translation-invariant phases, the number of remaining terms is of the order of the coordination number of the lattice.

The derivative in the last line of the previous equation evaluates to
\begin{eqnarray}
\fl\frac{\partial \Ket{\tilde\Phi_{abe}^s}\Bra{\tilde\Phi_{abe}^s}}{\partial\Phi_\nu^{r_1r_2}} =
i \Ket{\tilde\Phi_{abe}^s}\Bra{\tilde\Phi_{abe}^s} \odot \nonumber\\
\quad{}\odot \Big( \delta_{\nu,\nu(a,e)} \left[\delta_{\{r_1,r_2\},\{q_1,s\}} - \delta_{\{r_1,r_2\},\{q_1',s\}}\right] + \nonumber\\
\qquad{}+ \delta_{\nu,\nu(b,e)}\left[\delta_{\{r_1,r_2\},\{q_2,s\}} - \delta_{\{r_1,r_2\},\{q_2',s\}}\right] \Big)_{\mathbf{q},\mathbf{q}'\in\mathbb{S}^2}. \nonumber
\end{eqnarray}

\section*{References}
\addcontentsline{toc}{section}{References}

\bibliographystyle{simon2}
\bibliography{simon}

\newcommand{\etalchar}[1]{$^{#1}$}
\begin{thebibliography}{WVHC07}
\providecommand{\url}[1]{#1}
\providecommand{\urlprefix}{URL }
\expandafter\ifx\csname urlstyle\endcsname\relax
  \providecommand{\doi}[1]{doi:\discretionary{}{}{}#1}\else
  \providecommand{\doi}{doi:\discretionary{}{}{}\begingroup
  \urlstyle{rm}\Url}\fi
\providecommand{\eprint}[2][]{#2}
\providecommand{\arxiv}[1]{\href{http://www.arxiv.org/abs/#1}{ArXiv: #1}}
\providecommand{\arxivs}[2]{\href{http://www.arxiv.org/abs/#1}{ArXiv: #1 [#2]}}

\bibitem[ADG{\etalchar{+}}05]{ALPS}
F.~Alet, P.~Dayal, A.~Grzesik, A.~Honecker, M.~K\"orner, A.~L\"auchli, S.~R.
  Manmana, I.~P. McCulloch, F.~Michel, R.~M. Noack, G.~Schmid, U.~Schollw\"ock,
  F.~St\"ockli, S.~Todo, S.~Trebst, M.~Troyer, P.~Werner, S.~Wessel (ALPS
  collaboration).
\newblock \emph{The ALPS Project: Open source software for strongly correlated
  systems}.
\newblock J.~Phys. Soc. Jpn. (Suppl.) \textbf{74} (2005), 30.
\newblock \eprint{\arxiv{cond-mat/0410407}}.

\bibitem[AEPW02]{AreaLaw1}
K.~Audenaert, J.~Eisert, M.~B. Plenio, R.~F. Werner.
\newblock \emph{Entanglement properties of the harmonic chain}.
\newblock {}\href{http://dx.doi.org/10.1103/PhysRevA.66.042327}{Phys. Rev. A
  \textbf{66} (2002), 042327}.

\bibitem[Aki96]{Akima3D}
H.~Akima.
\newblock \emph{{Algorithm 761}: scattered-data surface fitting that has the
  accuracy of a cubic polynomial}.
\newblock {}\href{http://dx.doi.org/10.1145/232826.232856}{ACM Trans. Math.
  Softw. \textbf{22} (1996), 362}.

\bibitem[APD{\etalchar{+}}06]{WGSMinPRL}
S.~Anders, M.~B. Plenio, W.~D\"ur, F.~Verstraete, H.-J. Briegel.
\newblock \emph{Ground state approximation for strongly interacting systems in
  arbitrary dimension}.
\newblock {}\href{http://dx.doi.org/10.1103/PhysRevLett.97.107206}{Phys. Rev.
  Lett. \textbf{97} (2006), 107206}.

\bibitem[B{\etalchar{+}}]{SWIG}
D.~M. Beazley, et~al.
\newblock \emph{Simple Wrapper and Interface Generator (SWIG)}.
\newblock \url{http://www.swig.org}.

\bibitem[BCG05]{Python_for_linalg}
O.~Br\"oker, O.~Chinellato, R.~Geus.
\newblock \emph{Using Python for large scale linear algebra applications}.
\newblock {}\href{http://dx.doi.org/10.1016/j.future.2005.02.001}{Future Gener.
  Comput. Syst. \textbf{21} (2005), 969}.

\bibitem[BCS06]{entanglement-crit-2d}
T.~Barthel, M.-C. Chung, U.~Schollw\"{o}ck.
\newblock \emph{Entanglement scaling in critical two-dimensional fermionic and
  bosonic systems}.
\newblock {}\href{http://dx.doi.org/10.1103/PhysRevA.74.022329}{Phys. Rev. A
  \textbf{74} (2006), 022329}.

\bibitem[BLN95]{L-BFGS}
R.~H. Byrd, P.~Lu, J.~Nocedal.
\newblock \emph{A limited memory algorithm for bound constrained optimization}.
\newblock {}\href{http://dx.doi.org/10.1137/0916069}{SIAM J. Sci. Stat. Comp.
  \textbf{16} (1995), 1190}.

\bibitem[BM71]{BarouchMcCoy}
E.~Barouch, B.~M. McCoy.
\newblock \emph{Statistical mechanics of the XY model. II. Spin-correlation
  functions}.
\newblock {}\href{http://dx.doi.org/10.1103/PhysRevA.3.786}{Phys. Rev. A
  \textbf{3} (1971), 786}.

\bibitem[BR01]{cluster_states}
H.-J. Briegel, R.~Rau{\ss}endorf.
\newblock \emph{Persistent entanglement in arrays of interacting particles}.
\newblock {}\href{http://dx.doi.org/10.1103/PhysRevLett.86.910}{Phys. Rev.
  Lett. \textbf{86} (2001), 910}.

\bibitem[CEP07]{area_law_and_statistics}
M.~Cramer, J.~Eisert, M.~B. Plenio.
\newblock \emph{Statistics dependence of the entanglement entropy}.
\newblock {}\href{http://dx.doi.org/10.1103/PhysRevLett.98.220603}{Phys. Rev.
  Lett. \textbf{98} (2007), 220603}.

\bibitem[CEPD06]{area_law_harmonic_lattice}
M.~Cramer, J.~Eisert, M.~B. Plenio, J.~Dreissig.
\newblock \emph{Entanglement-area law for general bosonic harmonic lattice
  systems}.
\newblock {}\href{http://link.aps.org/abstract/PRA/v73/e012309}{Phys. Rev. A
  \textbf{73} (2006), 012309}.

\bibitem[CHDB05]{SpinGasPRL}
J.~Calsamiglia, L.~Hartmann, W.~D\"ur, H.-J. Briegel.
\newblock \emph{Spin gases: quantum entanglement driven by classical
  kinematics}.
\newblock {}\href{http://link.aps.org/abstract/PRL/v95/e180502}{Phys. Rev.
  Lett. \textbf{95} (2005), 180502}.

\bibitem[Cla04]{NatGrid}
F.~Clare.
\newblock \emph{NATGRID}.
\newblock Part of the \textit{NCAR Graphics} software library,
  \url{http://ngwww.ucar.edu/ngdoc/ng/ngmath/natgrid/intro.html} (2004).
\newblock [based on \textit {nngridr} by D. Watson (1994)].

\bibitem[DHH{\etalchar{+}}05]{spin_chain_long_range}
W.~D\"ur, L.~Hartmann, M.~Hein, M.~Lewenstein, H.-J. Briegel.
\newblock \emph{Entanglement in spin chains and lattices with long-range
  Ising-type interactions}.
\newblock {}\href{http://link.aps.org/abstract/PRL/v94/e097203}{Phys. Rev.
  Lett. \textbf{94} (2005), 097203}.

\bibitem[DKSV04]{time-evolution_DMRG_2}
A.~J. Daley, C.~Kollath, U.~Schollw\"{o}ck, G.~Vidal.
\newblock \emph{Time-dependent density-matrix renormalization-group using
  adaptive effective Hilbert spaces}.
\newblock {}\href{http://stacks.iop.org/1742-5468/2004/P04005}{J.~Stat. Mech.:
  Theor. Exp.  (2004), P04005}.
\newblock \eprint{\arxiv{cond-mat/0403313}}.

\bibitem[Eis06]{DMRG_Minimization_is_NP_complete}
J.~Eisert.
\newblock \emph{Computational difficulty of global variations in the density
  matrix renormalization group}.
\newblock {}\href{http://dx.doi.org/10.1103/PhysRevLett.97.260501}{Phys. Rev.
  Lett. \textbf{97} (2006), 260501}.

\bibitem[ELM93]{LoopQMC_first}
H.~G. Evertz, G.~Lana, M.~Marcu.
\newblock \emph{Cluster algorithm for vertex models}.
\newblock {}\href{http://dx.doi.org/10.1103/PhysRevLett.70.875}{Phys. Rev.
  Lett. \textbf{70} (1993), 875}.

\bibitem[EO06]{entanglement_scaling_time_evol}
J.~Eisert, T.~J. Osborne.
\newblock \emph{General entanglement scaling laws from time evolution}.
\newblock {}\href{http://link.aps.org/abstract/PRL/v97/e150404}{Phys. Rev.
  Lett. \textbf{97} (2006), 150404}.

\bibitem[FWGF89]{BH_first}
M.~P.~A. Fisher, P.~B. Weichman, G.~Grinstein, D.~S. Fisher.
\newblock \emph{Boson localization and the superfluid-insulator transition}.
\newblock {}\href{http://dx.doi.org/10.1103/PhysRevB.40.546}{Phys. Rev. B
  \textbf{40} (1989), 546}.

\bibitem[GME{\etalchar{+}}02]{SF_MI_trans_exp}
M.~Greiner, O.~Mandel, T.~Esslinger, T.~W. H\"ansch, I.~Bloch.
\newblock \emph{Quantum phase transition from a superfluid to a Mott insulator
  in a gas of ultracold atoms}.
\newblock {}\href{http://dx.doi.org/10.1038/415039a}{Nature \textbf{415}
  (2002), 39}.

\bibitem[GTJ{\etalchar{+}}03]{GSL}
M.~Galassi, J.~Theiler, G.~Jungman, B.~Gough, J.~Davies, R.~Priedhorsky,
  M.~Booth, F.~Rossi.
\newblock \emph{The GNU Scientific Library (GSL), version 1.3}.
\newblock \url{http://www.gnu.org/software/gsl/} (2003).
\newblock Software.

\bibitem[HCDB05]{spin_gases}
L.~Hartmann, J.~Calsamiglia, W.~D\"ur, H.-J. Briegel.
\newblock \emph{Spin gases as microscopic models for non-Markovian
  decoherence}.
\newblock {}\href{http://link.aps.org/abstract/PRA/v72/e052107}{Phys. Rev. A
  \textbf{72} (2005), 052107}.

\bibitem[HDE{\etalchar{+}}06]{marcs_review}
M.~Hein, W.~D\"ur, J.~Eisert, R.~Rau{\ss}endorf, M.~{\MakeUppercase{v}an
  den}~Nest, H.-J. Briegel.
\newblock \emph{Entanglement in graph states and its applications}.
\newblock In: G.~Casati, D.~Shepelyansky, P.~Zoller, G.~Benenti (Eds.),
  \emph{Quantum computers, algorithms and chaos}, \emph{International School of
  Physics Enrico Fermi}, vol. 162. IOS Press, Amsterdam (2006).
\newblock \eprint{\arxiv{quant-ph/0602096}}.

\bibitem[Hen84]{XY_2D_PhaseDiagram}
M.~Henkel.
\newblock \emph{Statistical mechanics of the 2D quantum XY model in a
  transverse field}.
\newblock {}\href{http://dx.doi.org/10.1088/0305-4470/17/14/013}{J.~Phys. A:
  Math. Gen. \textbf{17} (1984), L795}.

\bibitem[IJK05]{entanglement_XY_analytical_B}
A.~R. Its, B.-Q. Jin, V.~E. Korepin.
\newblock \emph{Entanglement in the XY spin chain}.
\newblock {}\href{http://stacks.iop.org/0305-4470/38/2975}{J.~Phys. A: Math.
  Gen. \textbf{38} (2005), 2975}.

\bibitem[JBC{\etalchar{+}}98]{BH_in_optical_lattices}
D.~Jaksch, C.~Bruder, J.~I. Cirac, C.~W. Gardiner, P.~Zoller.
\newblock \emph{Cold bosonic atoms in optical lattices}.
\newblock {}\href{http://dx.doi.org/10.1103/PhysRevLett.81.3108}{Phys. Rev.
  Lett. \textbf{81} (1998), 3108}.

\bibitem[JK04]{entanglement_XX_analytical}
B.-Q. Jin, V.~E. Korepin.
\newblock \emph{Quantum spin chain, Toeplitz determinants and the
  Fisher--Hartwig conjecture}.
\newblock
  {}\href{http://dx.doi.org/10.1023/B:JOSS.0000037230.37166.42}{J.~Stat. Phys.
  \textbf{116} (2004), 79}.

\bibitem[Kat62]{KatsuraXYZ}
S.~Katsura.
\newblock \emph{Statistical mechanics of the anisotropic linear Heisenberg
  model}.
\newblock {}\href{http://dx.doi.org/10.1103/PhysRev.127.1508}{Phys. Rev.
  \textbf{127} (1962), 1508}.

\bibitem[KGV83]{simulated_annealing}
S.~Kirkpatrick, C.~D. Gelatt, M.~P. Vecchi.
\newblock \emph{Optimization by simulated annealing}.
\newblock {}\href{http://dx.doi.org/10.1126/science.220.4598.671}{Science
  \textbf{220} (1983), 671}.

\bibitem[KH04]{QMC_Lattice_Review_04}
N.~Kawashima, K.~Harada.
\newblock \emph{Recent developments of world-line Monte Carlo methods}.
\newblock {}\href{http://dx.doi.org/10.1143/JPSJ.73.1379}{J.~Phys. Soc. Jpn.
  \textbf{73} (2004), 1379}.

\bibitem[KM05]{entanglement_crit_1D}
J.~P. Keating, F.~Mezzadri.
\newblock \emph{Entanglement in quantum spin chains, symmetry classes of random
  matrices, and conformal field theory}.
\newblock {}\href{http://dx.doi.org/10.1103/PhysRevLett.94.050501}{Phys. Rev.
  Lett. \textbf{94} (2005), 050501}.

\bibitem[Kor04]{entanglement_crit_1D_gapless}
V.~E. Korepin.
\newblock \emph{Universality of entropy scaling in one dimensional gapless
  models}.
\newblock {}\href{http://dx.doi.org/10.1103/PhysRevLett.92.096402}{Phys. Rev.
  Lett. \textbf{92} (2004), 096402}.

\bibitem[Lat07]{Latorre_on_MPS_and_QuantSim}
J.~I. Latorre.
\newblock \emph{Entanglement entropy and the simulation of quantum mechanics}.
\newblock {}\href{http://dx.doi.org/10.1088/1751-8113/40/25/S13}{J.~Phys. A:
  Math. Theor. \textbf{40} (2007), 6689}.

\bibitem[LLRV05]{EnriqueXY_EntropyPlot}
J.~I. Latorre, C.~A. Lutken, E.~Rico, G.~Vidal.
\newblock \emph{Fine-grained entanglement loss along renormalization-group
  flows}.
\newblock {}\href{http://dx.doi.org/10.1103/PhysRevA.71.034301}{Phys. Rev. A
  \textbf{71} (2005), 034301}.

\bibitem[LMSY96]{ARPACK}
R.~Lehoucq, K.~Maschhoff, D.~Sorensen, C.~Yang.
\newblock \emph{ARPACK}.
\newblock \url{http://www.caam.rice.edu/software/ARPACK/} (1996).

\bibitem[LRV04]{EnriqueGroundSpin}
J.~I. Latorre, E.~Rico, G.~Vidal.
\newblock \emph{Ground state entanglement in quantum spin chains}.
\newblock Quant. Inf. Comp. \textbf{4} (2004), 48.

\bibitem[LSA{\etalchar{+}}07]{Review_Optical_Lattices}
M.~Lewenstein, A.~Sanpera, V.~Ahufinger, B.~Damski, A.~{Sen De}, U.~Sen.
\newblock \emph{Ultracold atomic gases in optical lattices: Mimicking condensed
  matter physics and beyond}.
\newblock {}\href{http://dx.doi.org/10.1080/00018730701223200}{Adv. Phys.
  \textbf{56} (2007), 243}.
\newblock \eprint{\arxiv{cond-mat/0606771}}.

\bibitem[LXW03]{tDMRG_first}
H.~G. Luo, T.~Xiang, X.~Q. Wang.
\newblock \emph{Comment on ``Time-dependent density-matrix renormalization
  group: a systematic method for the study of quantum many-body
  out-of-equilibrium systems''}.
\newblock {}\href{http://dx.doi.org/10.1103/PhysRevLett.91.049701}{Phys. Rev.
  Lett. \textbf{91} (2003), 049701}.

\bibitem[MMN05]{tDMRG}
S.~R. Manmana, A.~Muramatsu, R.~M. Noack.
\newblock \emph{Time evolution of one-dimensional quantum many body systems}.
\newblock In: A.~Avella, F.~Mancini (Eds.), \emph{Lectures on the physics of
  highly correlated electron systems IX: 9th training course in the physics of
  correlated electron systems and high-$T_c$ superconductors}, \emph{AIP Conf.
  Proc.}, vol. 789, pp. 269--278. American Institute of Physics (2005).
\newblock \href{http://link.aip.org/link/?APC/789/269/1}{[web link]}.

\bibitem[MT94]{linesearchMoreThuente}
J.~J. Mor\'e, D.~J. Thuente.
\newblock \emph{Line search algorithms with guaranteed sufficient decrease}.
\newblock {}\href{http://dx.doi.org/10.1145/192115.192132}{ACM Trans. Math.
  Softw.  (1994), 286}.

\bibitem[MVC07]{PEPS_for_BH}
V.~Murg, F.~Verstraete, J.~I. Cirac.
\newblock \emph{Variational study of hard-core bosons in a two-dimensional
  optical lattice using projected entangled pair states}.
\newblock {}\href{http://link.aps.org/abstract/PRA/v75/e033605}{Phys. Rev. A
  \textbf{75} (2007), 033605}.

\bibitem[NH95]{deriv_matr_exp}
I.~Najfeld, T.~F. Havel.
\newblock \emph{Derivatives of the matrix exponential and their computation}.
\newblock {}\href{http://dx.doi.org/doi:10.1006/aama.1995.1017}{Advanc. Appl.
  Math. \textbf{16} (1995), 321}.

\bibitem[NM64]{NelderMead}
J.~A. Nelder, R.~Mead.
\newblock \emph{A simplex method for function minimization}.
\newblock Computer J. \textbf{7} (1964), 308.

\bibitem[NMDB06]{UniversalResourcePRL}
M.~{\MakeUppercase{v}an den}~Nest, A.~Miyake, W.~D\"ur, H.~J. Briegel.
\newblock \emph{Universal resources for measurement-based quantum computation}.
\newblock {}\href{http://link.aps.org/abstract/PRL/v97/e150504}{Phys. Rev.
  Lett. \textbf{97} (2006), 150504}.

\bibitem[Num]{NumPyURL}
\emph{NumPy}.
\newblock \url{http://numpy.scipy.org/}.

\bibitem[NW99]{NocedalWright}
J.~Nocedal, S.~J. Wright.
\newblock \emph{Numerical optimization}.
\newblock Springer (1999).

\bibitem[OAFF02]{entanglement_at_QPT}
A.~Osterloh, L.~Amico, G.~Falci, R.~Fazio.
\newblock \emph{Scaling of entanglement close to a quantum phase transition}.
\newblock {}\href{http://dx.doi.org/10.1038/416608a}{Nature \textbf{416}
  (2002), 608}.

\bibitem[Oli06]{NumPyGuide}
T.~E. Oliphant.
\newblock \emph{Guide to NumPy}.
\newblock Trelgol, \url{http://www.trelgol.com} (2006).

\bibitem[ON02]{entanglement_and_DMRG}
T.~J. Osborne, M.~A. Nielsen.
\newblock \emph{Entanglement, quantum phase transitions, and density matrix
  renormalization}.
\newblock {}\href{http://dx.doi.org/10.1038/416608a}{Quant. Inf. Proc.
  \textbf{1} (2002), 45}.

\bibitem[{\"O}R95]{DMRG-MPS-PRL}
S.~{\"O}stlund, S.~Rommer.
\newblock \emph{Thermodynamic limit of density matrix renormalization}.
\newblock {}\href{http://dx.doi.org/10.1103/PhysRevLett.75.3537}{Phys. Rev.
  Lett. \textbf{75} (1995), 3537}.

\bibitem[PEDC05]{AreaLaw2}
M.~B. Plenio, J.~Eisert, J.~Dreissig, M.~Cramer.
\newblock \emph{Entropy, entanglement, and area: Analytical results for
  harmonic lattice systems}.
\newblock {}\href{http://link.aps.org/abstract/PRL/v94/e060503}{Phys. Rev.
  Lett. \textbf{94} (2005), 060503}.

\bibitem[Pes04]{entanglement_XY_analytical_A}
I.~Peschel.
\newblock \emph{On the entanglement entropy for an XY spin chain}.
\newblock {}\href{http://stacks.iop.org/1742-5468/2004/P12005}{J.~Stat. Mech.:
  Th. Exp. \textbf{2004} (2004), P12005}.

\bibitem[Pet]{f2py}
P.~Peterson.
\newblock \emph{F2PY: Fortran to Python interface generator}.
\newblock \url{http://cens.ioc.ee/projects/f2py2e/}.

\bibitem[Pfe69]{PfeutyIsing}
P.~Pfeuty.
\newblock \emph{The one-dimensional Ising model with a transverse field}.
\newblock {}\href{http://dx.doi.org/10.1016/0003-4916(70)90270-8}{Annals of
  Physics \textbf{57} (1969), 79}.

\bibitem[PlM]{deriv_inv_matr}
\emph{Derivative of inverse matrix}.
\newblock PlanetMath.Org (a web encyclop{\ae}dia),
  \url{http://planetmath.org/encyclopedia/DerivativeOfInverseMatrix.html}
  (2006).
\newblock Version 3.

\bibitem[Pow64]{Powell_minimization}
M.~J.~D. Powell.
\newblock \emph{An efficient method for finding the minimum of a function of
  several variables without calculating derivatives}.
\newblock Computer J. \textbf{7} (1964), 155 .

\bibitem[PRB06]{CakeStructure}
G.~Pupillo, A.~M. Rey, G.~G. Batrouni.
\newblock \emph{Bragg spectroscopy of trapped one-dimensional strongly
  interacting bosons in optical lattices: Probing the cake structure}.
\newblock {}\href{http://link.aps.org/abstract/PRA/v74/e013601}{Phys. Rev. A
  \textbf{74} (2006), 013601}.

\bibitem[PST98]{WormQMC_first}
N.~V. Prokof'ev, B.~V. Svistunov, I.~S. Tupitsyn.
\newblock \emph{Exact, complete, and universal continuous-time worldline {Monte
  Carlo} approach to the statistics of discrete quantum systems}.
\newblock Zh. \'Eksp. Teor. Fiz. \textbf{114} (1998), 570.
\newblock Also: \href{http://dx.doi.org/10.1134/1.558661}{J.\ Exp.\ Th.\ Phys.\
  \textbf{87} (1998), 310}.

\bibitem[Puz05]{CayleyTransform}
R.~Puzio.
\newblock \emph{Cayley's parameterization of orthogonal matrices}.
\newblock PlanetMath.Org (a web encyclop{\ae}dia),
  \url{http://planetmath.org/?op=getobj&from=objects&id=6535} (2005).
\newblock Version 12.

\bibitem[R{\etalchar{+}}]{PythonHL}
G.~van Rossum, et~al.
\newblock \emph{Python {\rm [a programming language]}}.
\newblock \url{http://www.python.org}.

\bibitem[RBB03]{Long1QC}
R.~Rau\ss{}endorf, D.~E. Browne, H.-J. Briegel.
\newblock \emph{Measurement-based quantum computation on cluster states}.
\newblock {}\href{http://dx.doi.org/10.1103/PhysRevA.68.022312}{Phys. Rev. A
  \textbf{68} (2003), 022312}.

\bibitem[R{\"O}97]{DMRG_uses_MPS}
S.~Rommer, S.~{\"O}stlund.
\newblock \emph{Class of ansatz wave functions for one-dimensional spin systems
  and their relation to the density matrix renormalization group}.
\newblock {}\href{http://dx.doi.org/10.1103/PhysRevB.55.2164}{Phys. Rev. B
  \textbf{55} (1997), 2164}.

\bibitem[RT87]{clustering_globopt}
A.~H.~G. {Rinnooy Kan}, G.~T. Timmer.
\newblock \emph{Stochastic global optimization methods. Part I: Clustering
  methods}.
\newblock {}\href{http://dx.doi.org/10.1007/BF02592070}{Mathematical
  Programming \textbf{39} (1987), 27}.

\bibitem[Sac99]{Sachdev_book}
S.~Sachdev.
\newblock \emph{Quantum phase transitions}.
\newblock Cambridge Univesity Press, Cambridge (1999).

\bibitem[Sch02]{review_twophase}
F.~Schoen.
\newblock \emph{Two-phase methods for global optimization}.
\newblock In: P.~M. Pardalos, H.~E. Romeijn (Eds.), \emph{Handbook of global
  optimization, Vol. 2}. Kluwer (2002).
\newblock \eprint{\arxiv{quant-ph/0602096}}.

\bibitem[Sch05]{Schollwoeck_Review}
U.~Schollw\"ock.
\newblock \emph{The density-matrix renormalization group}.
\newblock {}\href{http://link.aps.org/abstract/RMP/v77/p259}{Rev. Mod. Phys.
  \textbf{77} (2005), 259}.

\bibitem[Sib81]{Sibson_interpolation}
R.~Sibson.
\newblock \emph{A brief description of natural neighbor interpolation}.
\newblock In: V.~Barnett (Ed.), \emph{Interpreting multivariate data}, pp.
  21--36. John Wiley \& Sons, New York (1981).

\bibitem[SP97]{diffevol}
R.~Storn, K.~Price.
\newblock \emph{Differential evolution -- a simple and efficient heuristic for
  global optimization over continuous spaces}.
\newblock {}\href{http://dx.doi.org/10.1023/A:1008202821328}{J.~Global
  Optimization \textbf{11} (1997), 341}.

\bibitem[SWVC07]{MPS_and_Renyi}
N.~Schuch, M.~M. Wolf, F.~Verstraete, J.~I. Cirac.
\newblock \emph{Entropy scaling and simulability by matrix product states}.
\newblock \arxivs{0705.0292}{quant-ph} (2007).

\bibitem[TAH98]{Troyer_code}
M.~Troyer, B.~Ammon, E.~Heeb.
\newblock \emph{Parallel object oriented Monte Carlo simulations}.
\newblock In: D.~Caromel, R.~R. Oldehoeft, M.~Tholburn (Eds.), \emph{Computing
  in object-oriented parallel environments (Proc. ISCOPE 1998)}, \emph{Lecture
  Notes in Computer Science}, vol. 1505, p. 191. Springer, Berlin etc. (1998).
\newblock \href{http://www.springerlink.com/content/yd6dqqy8u3trtae9}{[web
  link]}.

\bibitem[Tak99]{1D_exact_models}
M.~Takahashi.
\newblock \emph{Thermodynamics of one-dimensional solvable models}.
\newblock Cambridge University Press, Cambridge, UK (1999).

\bibitem[TB97]{NumLinAlg}
L.~N. Trefethen, D.~Bau.
\newblock \emph{Numerical linear algebra}.
\newblock SIAM, Philadelphia (1997).

\bibitem[VC04a]{Verstr_Renorm}
F.~Verstraete, J.~I. Cirac.
\newblock \emph{Renormalization algorithms for quantum-many body systems in two
  and higher dimensions}.
\newblock \arxiv{cond-mat/0407066} (2004).

\bibitem[VC04b]{vbs_for_qc}
F.~Verstraete, J.~I. Cirac.
\newblock \emph{Valence-bond states for quantum computation}.
\newblock {}\href{http://dx.doi.org/10.1103/PhysRevA.70.060302}{Phys. Rev. A
  \textbf{70} (2004), 060302}.

\bibitem[VGC04]{DMRG_thermal}
F.~Verstraete, J.~J. {Garc\'ia-{}\MakeUppercase{R}ipoll}, J.~I. Cirac.
\newblock \emph{Matrix product density operators: simulation of
  finite-temperature and dissipative systems}.
\newblock {}\href{http://dx.doi.org/10.1103/PhysRevLett.93.207204}{Phys. Rev.
  Lett. \textbf{93} (2004), 207204}.

\bibitem[Vid04]{VidalTEvol}
G.~Vidal.
\newblock \emph{Efficient simulation of one-dimensional quantum many-body
  systems}.
\newblock {}\href{http://link.aps.org/abstract/PRL/v93/e040502}{Phys. Rev.
  Lett. \textbf{93} (2004), 040502}.

\bibitem[VLRK03]{Entanglement_in_Quant_Crit}
G.~Vidal, J.~I. Latorre, E.~Rico, A.~Kitaev.
\newblock \emph{Entanglement in quantum critical phenomena}.
\newblock {}\href{http://dx.doi.org/10.1103/PhysRevLett.90.227902}{Phys. Rev.
  Lett. \textbf{90} (2003), 227902}.

\bibitem[Voj03]{QPT_review_Vojta}
M.~Vojta.
\newblock \emph{Quantum phase transitions}.
\newblock {}\href{http://dx.doi.org/10.1088/0034-4885/66/12/R01}{Rep. Prog.
  Phys. \textbf{66} (2003), 2069}.

\bibitem[VPC04]{DMRG_seen_from_QIT}
F.~Verstraete, D.~Porras, J.~I. Cirac.
\newblock \emph{Density matrix renormalization group and periodic boundary
  conditions: a quantum information perspective}.
\newblock {}\href{http://link.aps.org/abstract/PRL/v93/e227205}{Phys. Rev.
  Lett. \textbf{93} (2004), 227205}.

\bibitem[VWPC06]{area_law_PEPS}
F.~Verstraete, M.~M. Wolf, D.~{Perez-Garcia}, J.~I. Cirac.
\newblock \emph{Criticality, the area law, and the computational power of
  projected entangled pair states}.
\newblock {}\href{http://link.aps.org/abstract/PRL/v96/e220601}{Phys. Rev.
  Lett. \textbf{96} (2006), 220601}.

\bibitem[Whi92]{White_DMRG_1}
S.~R. White.
\newblock \emph{Density matrix formulation for quantum renormalization groups}.
\newblock {}\href{http://dx.doi.org/10.1103/PhysRevLett.69.2863}{Phys. Rev.
  Lett. \textbf{69} (1992), 2863}.

\bibitem[Whi93]{White_DMRG_2}
S.~R. White.
\newblock \emph{Density-matrix algorithms for quantum renormalization groups}.
\newblock {}\href{http://dx.doi.org/10.1103/PhysRevB.48.10345}{Phys. Rev. B
  \textbf{48} (1993), 10345}.

\bibitem[Whi98]{Whites_DMRG_review}
S.~R. White.
\newblock \emph{Strongly correlated electron systems and the density
  renormalization group}.
\newblock {}\href{http://dx.doi.org/10.1016/S0370-1573(98)00010-6}{Phys. Rep.
  \textbf{301} (1998), 187}.

\bibitem[Wol06]{AreaLaw_Fermions}
M.~M. Wolf.
\newblock \emph{Violation of the entropic area law for fermions}.
\newblock {}\href{http://dx.doi.org/10.1103/PhysRevLett.96.010404}{Phys. Rev.
  Lett. \textbf{96} (2006), 010404}.

\bibitem[WS99]{review_basin_hopping}
D.~J. Wales, H.~A. Scheraga.
\newblock \emph{Global optimization of clusters, crystals, and biomolecules}.
\newblock {}\href{http://dx.doi.org/10.1126/science.285.5432.1368}{Science
  \textbf{285} (1999), 1368}.

\bibitem[WVHC07]{holographic_principle_in_spin_lattices}
M.~M. Wolf, F.~Verstraete, M.~B. Hastings, J.~I. Cirac.
\newblock \emph{Area laws in quantum systems: mutual information and
  correlations}.
\newblock \arxivs{0704.3906}{quant-ph} (2007).

\bibitem[ZBLN97]{L-BFGS-B}
C.~Zhu, R.~H. Byrd, P.~Lu, J.~Nocedal.
\newblock \emph{Algorithm 778: L-BFGS-B: Fortran subroutines for large-scale
  bound-constrained optimization}.
\newblock {}\href{http://dx.doi.org/10.1145/279232.279236}{ACM Trans. Math.
  Softw. \textbf{23} (1997), 550}.

\bibitem[ZV04]{DMRG_thermal_B}
M.~Zwolak, G.~Vidal.
\newblock \emph{Mixed-state dynamics in one-dimensional quantum lattice
  systems: a time-dependent superoperator renormalization algorithm}.
\newblock {}\href{http://link.aps.org/abstract/PRL/v93/e207205}{Phys. Rev.
  Lett. \textbf{93} (2004), 207205}.

\end{thebibliography}


\end{document}